\documentclass[aps,pra,twocolumn,showpacs,preprintnumbers,amsmath,amssymb,superscriptaddress,longbibliography]{revtex4-1}

\usepackage{mathrsfs}
\usepackage{amsfonts}
\usepackage{amsmath}
\usepackage{amssymb}
\usepackage{natbib}
\usepackage{bbm}
\usepackage{hyperref}
\usepackage[dvips]{graphicx}
\usepackage{soul} % For strike through for review
\usepackage{xcolor}

\begin{document}
	
	%\preprint{APS/123-QED}
	
	\title{Hybrid approximation approach to generation of atomic squeezing with quantum nondemolition measurements}% Force line breaks with \\
	%\title{Optical generation of spin squeezed state in a double-well trap}
	%\title{Optical generation of spin squeezed state using atom Bose-Einstein condensate in a double-well trap}
	%\thanks{A footnote to the article title}%
	
	\author{Ebubechukwu O. Ilo-Okeke}
	\affiliation{New York University Shanghai, 1555 Century Ave, Pudong, Shanghai 200122, China}  
	\affiliation{Department of Physics, School of Physical Sciences, Federal University of Technology, P. M. B. 1526, Owerri 460001, Nigeria}
	\affiliation{NYU-ECNU Institute of Physics at NYU Shanghai, 3663 Zhongshan Road North, Shanghai 200062, China}

	\author{Manikandan Kondappan}
	\affiliation{State Key Laboratory of Precision Spectroscopy, School of Physical and Material Sciences, East China Normal University, Shanghai 200062, China}
	\affiliation{New York University Shanghai, 1555 Century Ave, Pudong, Shanghai 200122, China}

	\author{Ping Chen}
	\affiliation{State Key Laboratory of Precision Spectroscopy, School of Physical and Material Sciences, East China Normal University, Shanghai 200062, China}
	\affiliation{New York University Shanghai, 1555 Century Ave, Pudong, Shanghai 200122, China}

	\author{Yuping Mao}
	\affiliation{State Key Laboratory of Precision Spectroscopy, School of Physical and Material Sciences, East China Normal University, Shanghai 200062, China}
	\affiliation{New York University Shanghai, 1555 Century Ave, Pudong, Shanghai 200122, China}  

	\author{Valentin Ivannikov}
	\affiliation{New York University Shanghai, 1555 Century Ave, Pudong, Shanghai 200122, China}
	\affiliation{NYU-ECNU Institute of Physics at NYU Shanghai, 3663 Zhongshan Road North, Shanghai 200062, China}

	\author{Tim Byrnes}
	\affiliation{State Key Laboratory of Precision Spectroscopy, School of Physical and Material Sciences,
		East China Normal University, Shanghai 200062, China}
	\affiliation{New York University Shanghai, 1555 Century Ave, Pudong, Shanghai 200122, China}
	\affiliation{NYU-ECNU Institute of Physics at NYU Shanghai, 3663 Zhongshan Road North, Shanghai 200062, China}
	\affiliation{Center for Quantum and Topological Systems (CQTS), NYUAD Research Institute, New York University Abu Dhabi, UAE}
	\affiliation{Department of Physics, New York University, New York, NY 10003, USA}
	
	\date{\today}% It is always \today, today,
	%  but any date may be explicitly specified
	
	%%%%%%%%%%%%%%%%%%%%%%%%%%%%%%%%%%%%%%%%%%%%%%%%%%%%%%%%%%%%%%%%%%
	%%%%%%%%%%%%%%%%%%%%%%%%%%%%%%%%%%%%%%%%%%%%%%%%%%%%%%%%%%%%%%%%%%
	\begin{abstract}
		We analyze a scheme that uses quantum nondemolition measurements to induce squeezing of a two-mode Bose-Einstein condensate in a double well trap. In a previous paper [Ilo-Okeke et al.  Phys. Rev. A \textbf{104}, 053324 (2021)], we introduced a model to solve exactly the wavefunction for all atom-light interaction times. Here, we perform approximations for the short interaction time regime, which is relevant for producing squeezing. Our approach uses a Holstein-Primakoff approximation for the atoms while we treat the light variables exactly. It allows us to show that the measurement induces correlations within the condensate, which manifest in the state of the condensate as a superposition of even parity states. In the long interaction time regime, our methods allow us to identify the mechanism for loss of squeezing correlation. We derive simple expressions for the variances of atomic spin variables conditioned on the measurement outcome. We find that the results agree with the exact solution in the short interaction time regime. Additionally, we show that the expressions are the sum of the variances of the atoms and the measurement. Beyond the short interaction time regime, our scheme agrees qualitatively with the exact solution for the spin variable that couples to light.
	\end{abstract}
	
	%	The second equation tells us that a finite variance of the distribution
	%	W(q) yields an additional dispersion which is superimposed on the intrinsic
	%	quantum fluctuations given by Var(Q).
	
	%\pacs{}% PACS, the Physics and Astronomy
	% Classification Scheme.
	%\keywords{Suggested keywords}%Use showkeys class option if keyword
	%display desired
	\maketitle

	%%%==========================================================
	%%% Introduction
	%%%==========================================================
	\section{Introduction \label{sec:intro}}
	%%%==========================================================
	Squeezed states improve precision measurements beyond the standard quantum limit attainable by classical means. As squeezed states possess  entanglement~\cite{einstein1935}, they surpass the standard quantum limit set by the quantum noise of individual uncorrelated quantum particles. Consequently, squeezed states are increasingly used for enhanced measurement and detection in many areas of physics such as image reconstruction~\cite{brida2010}, magnetometry and electric-field sensing~\cite{brask2015,fan2015}, optical interferometry~\cite{dowling2015,schnabel2017},  gravitational wave detection~\cite{eberle2010,pitkin2011}, and used as a resource~\cite{julsgaard2001,kuzmich2004,esteve2008} for heralding entanglement in many-body spin systems. 
	
	Squeezed states~\cite{walls1983,kitagawa1993,wineland1994,ma2011,schnabel2017,pezze2018} are realized in many different physical systems via interactions that induce correlations between quantum particles. For instance, correlations and squeezing arise naturally from interactions found in the quantum system, such as the two-body interactions in atomic Bose-Einstein condensates~\cite{kitagawa1993,esteve2008,bohi2009} and ions~\cite{leibfried2004,ge2019}, or the long-range interactions of the Rydberg atoms~\cite{saffman2010,gil2014}. Additionally, 
	squeezing and correlations are induced by interactions between two different quantum systems. A typical example is the generation of optical squeezed states~\cite{slusher1985, wu1986, josse2003} using the interactions between light and barium borate crystal. 
	
	Quantum nondemolition (QND) measurements have been shown to be an effective way of producing squeezing in a variety of systems~\cite{unruh1979,braginsky1980,roch1992,grangier1998,takahashi1999,kuzmich2000,higbie2005,lecocq2015}. In atomic systems, for instance, the quantum states of light and atoms become entangled as a result of interactions between atoms and light. Upon the detection of the light, the entanglement between quantum states of atom and light is transferred to that among the individual atoms~\cite{higbie2005,meppelink2010,ilo-okeke2014,ilo-okeke2016}, manifesting itself as the squeezed states of the atoms~\cite{kitagawa1993}. Such states possessing squeezing are viewed as a resource in quantum metrology~\cite{pezze2018}, and quantum information~\cite{cerf2007,abdelrahman2014,pyrkov2014,byrnes2015,feng2021}. Stroboscopic monitoring has been used to squeeze the atomic spins~\cite{thorne1978,braginsky1980,vasilakis2015,moller2017}. In this method, a measurement is timed to coincide with the period of the oscillator. These ensure that the spin component of the atom coupled to light experiences a minimum disturbance due to measurement---an effect termed measurement back-action~\cite{thorne1978,braginsky1980}. 
	
	In a previous work (Ref.~\cite{ilo-okeke2021}), we proposed a scheme that does not depend on timed measurements to achieve minimal disturbances due to measurement while achieving squeezing of the atomic states. In this scheme, the continuous monitoring of the atoms induces random motion on the mean spin whose projection is confined to a plane perpendicular mean field direction. In addition, the photon detection constricts the motion of the mean spin along the spin variable that couples to light thereby resulting in  spin squeezing. This squeezing manifests as the shearing of the distribution in phase-space~\cite{ilo-okeke2021}.
	
	Squeezing correlations emerge in interacting many-particle systems where the interaction in the system is weak. It is a precursor to other forms of correlations found in the system. As interactions within the system become strong, the squeezing correlations give way to other forms of correlation~\cite{pezze2018,ilo-okeke2021}, such as the Schr{\"o}dinger-cat state~\cite{ilo-okeke2021}. The Holstein-Primakoff (HP) approximation~\cite{holstein1940} is used to study the dynamics of many-particle systems in the large particle number limit. These include correlations, like squeezing, generated by weak interactions. In atom-light systems, usually described within QND measurements, there are many studies~\cite{hald1999,duan2000,julsgaard2001,kuzmich2000,wang2003,madsen2004,hammerer2010,vasilakis2015} on the squeezing of atoms employing the HP approximation~\cite{holstein1940}. However, all these studies to date focus on using the system's variables, such as spins or atom number, in showing the squeezing of atoms. To date, within the HP approximation, there are no works that give evidence of the squeezing correlation in atom systems using their quantum state as has been done for light~\cite{barnett1997,scully1997,walls2008}. In this paper, for the first time, we derive the quantum squeezed state of an atom oscillator interacting weakly with light within the HP approximation. The state we derive is an even parity state akin to what is obtained using light~\cite{barnett1997,scully1997,walls2008}.

	The rest of the paper is organized as follows. We begin by presenting the technique and a summary of the main result of this work in Sec.~\ref{sec:aim}. Next, we present the double-well oscillator model in Sec.~\ref{sec:model}. The hybrid approximation is presented in Sec.~\ref{sec:cv}. The hybrid approximation is used to analyze an initial state that is perfectly polarized in Sec.~\ref{sec:results}. In Sec.~\ref{sec:comparison}, we analyze an initial state with a small but finite population in the excited state using the hybrid approximation, and compare the results with the exact solution. We present the general feature of the conditional variance in the short time regime in Sec.~\ref{sec:discussion}, and finally present our summary and conclusions in Sec.~\ref{sec:summary}.

	%%%%%===========================================
	%%%%%Figure 1
	%%%%%===========================================
	\begin{figure}[t]
		\includegraphics[width=\columnwidth]{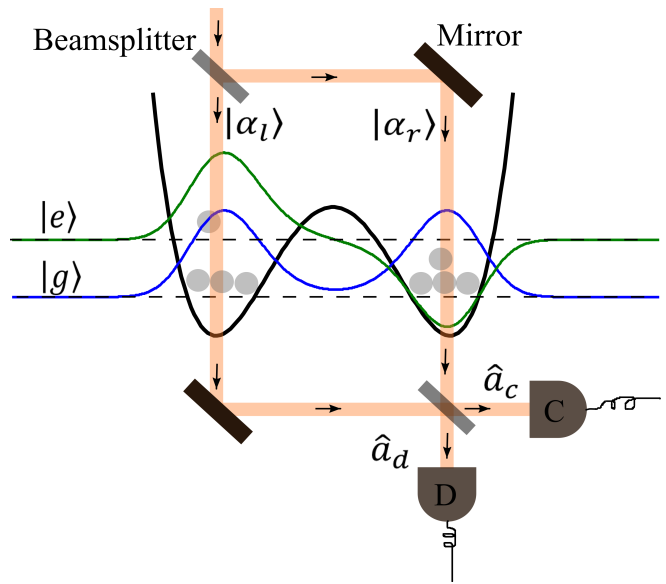}
		\caption{Atoms in a double-well trap are imaged using identical laser light in a Mach-Zehnder interferometer geometry. The available states in the trap are the symmetric ground state $\lvert g\rangle $ and the anti-symmetric first excited state $\lvert e\rangle$.} 
		\label{fig11}
	\end{figure}
	%%%%%===========================================

	%%%==========================================================
	\section{Main result of this paper\label{sec:aim}}
	%%%==========================================================
	We first give a basic introduction of the aim, main technique, and main result of this paper. As illustrated in Fig.~\ref{fig11}, the number of atoms in the system is macroscopically large $N\gg 1$, with virtually all of them found in the ground state of the trap, and an atomic Bose-Einstein condensate placed in the double well realizes this. Since the number of atoms in the ground state is large,  a classical number $\sqrt{N}$, it can be used to replace the mode operators of the ground state. Doing this gives rise to operators mathematically equivalent to photon operators. This step is the Holstein-Primakoff approximation, where the mode operators of the excited state map to the $\hat{X}$ and $\hat{P}$ of a harmonic oscillator. However, the light operators are not approximated. The resulting Hamiltonian is that of a displaced oscillator for the atomic operators because of the presence of light, $$H = -\frac{\hbar\Omega}{2} (\hat{X}^2 + \hat{P}^2)+ \sqrt{2N}\hbar g\hat{X} S_z,$$   where $\Omega$ is the tunneling frequency between the wells, $g$ is coupling frequency between atom and light, and $S_z$ is the relative photon number operator. The approach where we treat the mode operators of the double-well trap within the Holstein-Primakoff approximation while not approximating the photon operators is the key to deriving the squeezed state of an atom oscillator. We call this the ``hybrid approximation'', which allows us to calculate the quantum state of the atom oscillator. It also marks the point of departure of this work from previous studies~\cite{hald1999,duan2000,kuzmich2000,julsgaard2001,wang2003,madsen2004,hammerer2010,vasilakis2015} using the Holstein-Primakoff approximation~\cite{holstein1940}.

	Our calculation then proceeds as follows. Let $\lvert \psi_0\rangle$ be the initial state of the condensate, while $\lvert  \alpha_l\rangle,\, \lvert \alpha_r\rangle$ are the initial coherent state of light in the left and right arms of the interferometer, respectively. The combined state of the atom and light, $\lvert\psi_0,  \alpha_l, \alpha_r\rangle = \lvert\psi_0\rangle\otimes \lvert  \alpha_l\rangle\otimes \lvert \alpha_r\rangle$, evolves under the Hamiltonian as prescribed by Schr{\"o}dinger equation, $ e^{-iHt/\hbar}\lvert\psi_0, \alpha_l,\alpha_r\rangle$. At the beamsplitter, the light beams in the two arms of the interferometer interfere. The unitary operator that describes the action of the beam splitter is $e^{-i\pi S_x/2}$. Counting the photons in the light beams by the detectors, $c$ and $d$, collapses the quantum state of light. The collapse of the light state described by a projection operator $\lvert n_c,n_d\rangle\langle n_c,n_d\rvert$ gives the state of the condensate  conditioned on counting $n_c$ and $n_d$ photons at the detectors $c$ and $d$, respectively, as $$\lvert \psi_{n_c,n_d}\rangle = \langle n_c,n_d\rvert e^{-i\pi S_x/2} e^{-iHt/\hbar}\lvert\psi_0, \alpha_l,\alpha_r\rangle .$$
	At this point we assume that the relative photon number difference is smaller than the total number of photons collected at the detectors $c$ and $d$, $\lvert n_d - n_c\rvert\ll(n_c + n_d)$. It is the condition for which correlation is maximized. Additionally, to minimize the effect of measurement on the quantum state of the condensate containing $N$ atoms, we consider balanced detection: the average photon number in the arms of the interferometer is the same, $\alpha_r = \alpha_l$. At this condition, the effects due to the total photon number $n_c + n_d$ in a measurement outcome vanish, leaving off only those from the relative photon number difference, $n_d - n_c$.

	We state the main result that we find from the above calculation. For a perfectly polarized initial state $\lvert \psi_0\rangle$ of the condensate, where all the $N$ atoms are in the ground state of the double-well, the normalized state of the condensate conditioned on the counting $n_c$ and $n_d$ photons at the detectors $c$ and $d$, respectively, is
	\begin{equation*}
			\lvert\psi_{n_c,n_d}\rangle \propto  \sum_{n=0}^{N_\mathrm{max}}C_n\left(\lvert\alpha_r\rvert,gt,N\right)\lvert2 n\rangle,		
	\end{equation*}
	where $N_\mathrm{max} = N/2$ for $N$ even, and $N_\mathrm{max} = (N-1)/2$ for $N$ odd. The coefficients $C_{n}\left(\lvert\alpha\rvert,gt,N\right)$ depend on the average photon number at the input $\lvert\alpha_{r}\rvert^2$, the atom-light interaction strength $gt$ and the total number of atoms $N$, expressed in polynomial powers of $n$. Its explicit form is given in (\ref{eq:normalizedstate}). The above preceding equation is the main results of this paper.

	In the remainder of the paper, we give a detailed derivation of this state and use it to examine properties of the oscillator by calculating the averages such variance and expectation value of the oscillator quadratures $\hat{X}$ and $\hat{P}$, and compare the results to the exact solution. Additionally, we use the Q-function to highlight the various contributions from the initial state of the light and the detection of photons to the quantum state of the condensate. Given an imperfectly polarized state of the condensates, where a small but finite number of atoms are found in the excited state as shown in Fig.~\ref{fig11}, we showed that the normalized quantum state of the condensate conditioned on the detection of $n_c$ and $n_d$ is also an even parity state.

%	\textcolor{blue}{	
%	Squeezing is a  form of correlation found in many-particle system when the interaction in system is weak. It can be regarded as precursor to other forms of correlations that can be found in the system. As interaction within the system become strong, the squeezing correlations give way to other forms of correlation, like the Schr{\"o}dinger-cat state. The Holstein-Primakoff (HP) approximation is a well-developed technique for handling the evolution of many-body systems in the large particle number limit. It has been around for a while.
%	}

	%%%==========================================================
	\section{The double-well model\label{sec:model}}
	%%%==========================================================
	The model we consider has been previously described in Ref.~\cite{ilo-okeke2021} which we summarize briefly here. We consider an atomic Bose-Einstein condensate (BEC) confined in a double-well potential with the lowest energy states being the symmetric and antisymmetric states as shown in Fig.~\ref{fig11}. We further assume that fluctuation about the mean number density is small so that the resulting mean-field Hamiltonian can be written in second quantization as 
	\begin{equation}
		\label{eq:secondquantisedHamil}
		\hat{H} = \frac{1}{2}(\omega_e + \omega_g) \hat{N} + \hbar\Omega J_z,
	\end{equation}
	where $\hat{N} = \hat{b}^\dagger_g\hat{b}_g + \hat{b}^\dagger_e\hat{b}_e$ is the total atom number operator, $\hat{b}_{e,g} $ are the annihilation operators for the excited and ground states, respectively, and $\Omega = \omega_e - \omega_g$ is the particle tunneling frequency. The  spin operators are defined as
	\begin{equation}
		\label{eq:SpinOperators}
		\begin{split}
			J_x &= \frac{\hat{b}^\dagger_e\hat{b}_g + \hat{b}^\dagger_g\hat{b}_e}{2},\\
			J_y &= -i\frac{\hat{b}^\dagger_e\hat{b}_g - \hat{b}^\dagger_g\hat{b}_e}{2},\\
			J_z &= \frac{\hat{b}^\dagger_e\hat{b}_e - \hat{b}^\dagger_g\hat{b}_g}{2}.
		\end{split}
	\end{equation}
	
	The atoms in the double-well potential interact with laser light that is detuned from atomic resonance transition. Such interactions are described by the QND Hamiltonian~\cite{takahashi1999,julsgaard2001,kuzmich2004,ilo-okeke2014,behbood2014}. Hence the total Hamiltonian of the atom-light system is
	\begin{equation}
		\label{eq:FullHamiltonian}
		\hat{H} =\hbar\Omega J_z + 2\hbar g J_x S_z, 
	\end{equation} 	
	where $g$ is the atom-light coupling strength, and we have neglected the constant energy term that contributes a global phase shift in the states of atom and light in (\ref{eq:FullHamiltonian}). The dependence of the atom-light interactions on the spins arises from writing the sum on atom-light interactions in both wells in terms of the relative photon number $S_z = (\hat{a}^\dagger_l\hat{a}_l - \hat{a}^\dagger_r\hat{a}_r)/2$.
	
	The initial state of $N$ atoms in the BEC is taken to be a spin coherent state formed from the superposition of the excited and ground state as~\cite{byrnes2021}
	\begin{equation}
		\label{eq:AtomCoherentState}
		\lvert\alpha,\beta\rangle\rangle = \frac{(\alpha b^\dagger_e + \beta b^\dagger_g)^N}{\sqrt{N!}} \lvert \mathrm{vac}\rangle,
	\end{equation} 
	where $\lvert \mathrm{vac}\rangle$ is the vacuum state, a state with no atoms or photons. The amplitudes $\alpha$ and $\beta$ give the fractional population of atoms in the excited and ground state, respectively. Similarly, the quantum state of the laser light in the two arms of the interferometer is a coherent state of the form $\lvert \alpha_l\rangle\otimes\lvert \alpha_r\rangle$, where
	\begin{equation}
		\label{eq:LightCoherentState}
		\lvert\alpha_s\rangle = e^{-\frac{|\alpha_s|^2}{2}}e^{\alpha_s \hat{a}^\dagger_s}\lvert \mathrm{vac}\rangle,
	\end{equation}
	and $s = r,l$. 
	
	Given the initial states of atoms (\ref{eq:AtomCoherentState}) and light (\ref{eq:LightCoherentState}), the combined quantum state of atom and light at any other time is governed by  (\ref{eq:FullHamiltonian}) as 	
	\begin{equation}
		\label{eq:StateEvolution}
		\lvert\Psi\rangle = e^{-i\frac{\hat{H}t}{\hbar}}\lvert\alpha,\beta\rangle\rangle\otimes\lvert\alpha_l\rangle\otimes\lvert\alpha_r\rangle,
	\end{equation} 
	and is a solution of the Schr\"odinger equation
	\begin{equation}
		\label{eq:papc29}
		i\hbar\frac{d\lvert\Psi\rangle}{dt} = 	\hat{H}\lvert\Psi\rangle.
	\end{equation}

	This entangles the atoms and the light.  After this, the light in the Mach-Zehnder interferometer are interfered and measured.  We previously derived an exact solution of this valid for all interaction times~\cite{ilo-okeke2021}. Although exact, the resulting expressions are typically complex due to the many-body entangled wavefunction that results from (\ref{eq:StateEvolution}).  Our aim in this paper will be to obtain some simple formula for the relevant quantities, so that one can obtain a greater understanding of the important features and parameters governing the behavior of a quantum system.   For instance, in the case at hand, it will be interesting to understand how short the evolution time could be to sustain squeezing. One could also be interested in describing the squeezing effect in terms of parameters of the atom and light in the short time regime. Such understanding offers insight into the control and manipulation of squeezed states of atoms.

	%%%==========================================================
	\section{The large \emph{N }limit: The hybrid approximation \label{sec:cv}}
	%%%========================================================== 
	In this section, we describe the quantum state of an atom within the Holstein-Primakoff approximation. This is motivated by the fact that at short atom-light interaction times, the ground state population is nearly undepleted $\langle J_z\rangle \sim N/2$, even in the presence of a small dephasing parameter, $\gamma <g$ (see Figs. 5 and 8 of Ref.~\cite{ilo-okeke2021}). The results of this section will be used to analyze various physical quantities for two different initial states of the atoms, a perfect and an imperfectly spin polarized atoms, in  Secs. \ref{sec:results} and \ref{sec:comparison}.
	
	Because of the large number $N$ of atoms $N\gg1$, and given that at short times, the atoms are nearly in the ground state of the double-well trap, the Hamiltonian (\ref{eq:FullHamiltonian}) within the Holstein-Primakoff approximation, reduces to
	\begin{equation}
		\label{eq:strn:cv01}
		H \approx -\frac{\hbar\Omega}{2}\left(\hat{X}^2 + \hat{P}^2\right) +  \sqrt{2N}\hbar g \hat{X}S_z,
	\end{equation}
	where $\hat{X} = J_x/\sqrt{N/2}$, $\hat{P} = J_y/\sqrt{N/2}$ and a term proportional to the total atom number $N$ is neglected in writing (\ref{eq:strn:cv01}). The operators $\hat{X}$ and $\hat{P}$ in terms of the harmonic oscillator operators are respectively
	\begin{equation}
		\label{eq:strn:cv02}
		\hat{X} = \frac{\hat{b}^\dagger_e + \hat{b}_e}{\sqrt{2}},\qquad \hat{P} = i\frac{\hat{b}^\dagger_e - \hat{b}_e}{\sqrt{2}},
	\end{equation}
	and satisfies commutation relation	$[\hat{X},\hat{P}] = i$.
	
	Given an initial state of the atoms, $\lvert\psi_0\rangle$, and in the presence of atom-light interactions $g \neq 0$, the initial state of atom-light system, $\lvert \psi_0,\alpha_l,\alpha_r\rangle = \lvert \psi_0\rangle\otimes\lvert\alpha_l\rangle\otimes\alpha_r\rangle $, evolves under the Hamiltonian (\ref{eq:strn:cv01}) as 
	\begin{equation}
		\label{eq:strn:cv04}
		\lvert\psi(t)\rangle = e^{i\Omega t\left[\frac{\hat{X}^2 + \hat{P}^2}{2}-\sqrt{2}\kappa\hat{X}(\hat{a}^\dagger_l\hat{a}_l -  \hat{a}^\dagger_r\hat{a}_r)\right] }\lvert \psi_0,\alpha_l,\alpha_r\rangle,
	\end{equation}
	where $\kappa = g\sqrt{N}/(2\Omega)$. This may also be written as
	\begin{align}
		\label{eq:strn:cv05}
		\lvert\psi(t)\rangle & = e^{-\frac{\lvert \alpha_l\rvert^2 + \lvert \alpha_r\rvert^2}{2}}\sum_{n_l=0}^\infty\sum_{n_r = 0}^\infty e^{i\Omega t\left[\frac{\hat{X}^2 + \hat{P}^2}{2} -\sqrt{2} \kappa (n_l-n_r) \hat{X}\right] }\nonumber\\
		& \times\frac{\left(\alpha_l\right)^{n_l}}{n_l!}(\hat{a}^\dagger_l)^{n_l}\frac{\left(\alpha_r\right)^{n_r}}{n_r!}(\hat{a}^\dagger_r)^{n_r} \lvert \psi_0\rangle\lvert \mathrm{vac}\rangle,
	\end{align}
	where $n_l$ and $n_r$ are the number of photons in the left and right arms of the interferometer, respectively.	It is readily seen that the phase of the photons contains information about the atomic BEC. This information carried by the photons is accessed by interfering the photons at the beamsplitter as shown in Fig.~\ref{fig11}. The interference of photons performed by beamsplitter corresponds to the following transformation 
	\begin{equation}
		\label{eq:OperatorTransformation}
		\hat{a}^\dagger_l = \frac{\hat{a}^\dagger_c + i\hat{a}^\dagger_d}{\sqrt{2}},\quad \hat{a}^\dagger_r =\frac{i\hat{a}^\dagger_c + \hat{a}^\dagger_d}{\sqrt{2}}.
	\end{equation} 
	Hence, the wave function immediately after recombination becomes
	\begin{align}
		\label{eq:strn:cv06}
		\lvert\psi_\mathrm{BS}(t) \rangle & =  e^{-\frac{\lvert \alpha_l\rvert^2}{2}-\frac{\lvert \alpha_r\rvert^2}{2}}\sum_{n_l=0}^{\infty}\sum_{n_r=0}^{\infty}e^{i\Omega t\left[\frac{\hat{X}^2 + \hat{P}^2}{2} -\sqrt{2} \kappa (n_l-n_r) \hat{X}\right]  }\nonumber\\
		&\times\frac{\left(\alpha_l\right)^{n_l}}{n_l!}
		\left(\frac{\hat{a}^\dagger_c + i\hat{a}^\dagger_d}{\sqrt{2}}\right)^{n_l} \frac{\left(\alpha_r\right)^{n_r}}{n_r!}\left(\frac{i\hat{a}^\dagger_c +\hat{a}^\dagger_d}{\sqrt{2}}\right)^{n_r}\nonumber \\
		&\times\lvert\psi_0\rangle\lvert \mathrm{vac}\rangle.
	\end{align}

	The state~(\ref{eq:strn:cv06}) that emerges after the interference of photons on the beamsplitter is a superposition of the states of atoms and light.  Counting $n_c$ and $n_d$ photons at the detector collapses the states of light. Such a measurement is described by the projection operator $\mathcal{P} = \rvert n_c, n_d \rangle\langle n_c, n_d\lvert$. The state of atoms (not normalized) $\lvert\psi_{n_c,n_d}\rangle = \mathcal{P}\rvert \psi_\mathrm{BS}(t)\rangle$ conditioned on photon detection is 
	\begin{align}
		\label{eq:conditionalstate}
		\lvert\tilde{\psi}_{n_c,n_d}\rangle = \frac{e^{-\frac{\lvert\alpha_l\rvert^2 +\lvert\alpha_r\rvert^2}{2}}}{\sqrt{n_c!} \sqrt{n_d!}}\left(\frac{\alpha_r}{\sqrt{2}}\right)^{n_c +n_d} \sum_{n_l = 0}^\infty e^{in_l\phi}\nonumber\\
		\times  I(n_c, n_d, n_l)e^{i\Omega t\left[\frac{\hat{X}^2 + \hat{P}^2}{2} + \sqrt{2}\kappa(n_c+n_d-2n_l)\hat{X}\right]} \lvert \psi_0\rangle,
	\end{align}
	where $\phi = \arg(\alpha_l) - \arg(\alpha_r)$, and the Fourier coefficient $I(n_c,n_d,n_l)$ is defined as 
	\begin{align}
		\label{eq:strn:cv09a}
		I(n_l,n_c,n_d) &= \frac{1}{2\pi}\int_{0}^{2\pi}d\phi\, 	e^{-i\,n_l\phi}\left( 1 + i\frac{|\alpha_l|}{|\alpha_r|}e^{i\phi}\right)^{n_c}\times\nonumber\\ &\left(i + \frac{|\alpha_l|}{|\alpha_r|}e^{i\phi} \right)^{n_d},
	\end{align}
	see Appendix \ref{sec:sec:FourierCoeff} for details. For the analysis of the Fourier coefficient, the interval $\left[0,2\pi\right]$ is convenient when the largest contribution of the integrand comes from points around $\phi \approx \pi$, whereas the interval will be shifted to $\left[-\pi,\pi\right]$ when the largest contribution of the integrand comes from points around $\phi \approx 0$.

	%%%==========================================================
	%%% 
	%%%==========================================================
	\section{Perfect initial spin polarization\label{sec:results}}
	%%%========================================================== 
	In this section, we derive the quantum state of the atoms after detection using~(\ref{eq:conditionalstate}). This allows us to establish the squeezing correlations induced by the measurement. Secondly, it allows us to estimate how short the measurement should be, that is the region of validity of Holstein-Primakoff approximation performed on the atoms. We then solve exactly the state of atom and light using the Schr\"odinger equation~(\ref{eq:papc29}), followed by the interference of the photons on a beamsplitter as shown in Fig.~\ref{fig11}. The exact wave function conditioned on the detection of photons is used to calculate the relevant averages of the atomic operators. We derive expressions for the mean and variance of the atomic variables under the hybrid approximation and compare them with the exact solution. Additionally, we visualize the evolution of the conditional squeezing in phase-space using the Husimi \emph{Q}-distribution.		
	
	%%===========================================
	%%==========================================================
	\subsection{State Of Condensate Conditioned On Photon Detection\label{sec:sec: probden}}
	%%==========================================================
	We consider the case where all the atoms are initially in the ground state of the double-well trap. The fractional probability amplitudes become $\alpha =0$ and $\beta = 1$. In this case, the initial state (\ref{eq:AtomCoherentState}) becomes
	\begin{equation}
		\label{eq:strn:cv03}
		\lvert\psi_0\rangle = \lvert 0,1\rangle\rangle = \lvert 0, N\rangle \equiv \lvert0\rangle,
	\end{equation}
	where the state $\lvert 0, N\rangle$ is a Fock state defined as 
	\begin{equation}
		\lvert 0, N\rangle = \frac{\left(b^\dagger_g\right)^N}{\sqrt{N!}}\lvert\mathrm{vac}\rangle.
	\end{equation}
	Equation~(\ref{eq:strn:cv03}) is the vacuum state of the excited state of the double-well trap. This state describes a perfectly polarized state along \emph{z}. It is easy to see from (\ref{eq:strn:cv02}) that although the atoms in the excited state execute harmonic motion, on average, the number of atoms in the excited state at any time for $g=0$ is zero, $\langle 0 \lvert \hat{X}\rvert 0\rangle = 0$. These point to the type of motion executed by the atoms is mostly due to fluctuation. Hence, the variance is  $(\Delta X)^2 = 1/2$. In the presence of light $g\neq 0$, the initial state $\lvert\psi_0\rangle$ evolves as prescribed by (\ref{eq:strn:cv04}), and is modified further by the detection of photons. 
	
	The unnormalized state of the atoms after detection of photons is given by~(\ref{eq:conditionalstate}). The operator $e^{i\Omega t\left[\frac{\hat{X}^2 + \hat{P}^2}{2} + \sqrt{2}\kappa(n_c+n_d-2n_l)\hat{X}\right]}$ in (\ref{eq:conditionalstate}) can be written as a product of operators $e^{ip(\theta)\hat{b}^\dagger_e } 	e^{iq(\theta)\hat{b}^\dagger_e \hat{b}_e} e^{ir(\theta)\hat{b}_e }e^{is(\theta)}$, see~(\ref{eq:dis03}). Of these only the terms having $s(\theta)$ and $\hat{b}^\dagger_e$ in the exponent have an effect on the initial state (\ref{eq:strn:cv03}) --- a state with zero spin variance along the \emph{z}-axis. The measurement has two effects on the state  $\lvert0\rangle$. It creates a state with finite width via the term  containing $s(\theta)$ from the initial state. Secondly, the term having $\hat{b}^\dagger_e$ then introduces correlations when the sum over all the input photon states $n_l$ is performed~(\ref{eq:conditionalstate}). 
	
	To see this, we evaluate the probability amplitude explicitly for the initial state (\ref{eq:strn:cv03}). Detailed calculations are provided in Appendices~\ref{sec:sec:FourierCoeff} and \ref{sec:probilityamplitude}, we show the final results here. 
	To minimize the effect of measurement, we examine the case where the measurement is balanced $\alpha_{l} = \alpha_{r}$. At this condition, the effects due to the total photon number $n_c + n_d$ is suppressed, leaving off those only from the relative photon number difference, $n_d - n_c$.
		
	Using the completeness relation for the atomic state $\sum_n\lvert n\rangle\langle n\rvert = \hat{\mathbbm{1}}$, the state of the atoms conditioned on detection of $n_c$, $n_d$ photons is
	\begin{align*}
		\lvert\tilde{\psi}_{n_c,n_d}\rangle &\propto \sum_{n = 0}^{N}\sum_{k=0}^{\frac{n}{2}} \frac{n!}{k!(n-2k)!} \left(\frac{2\lvert\alpha_r\rvert^2}{1+ 4\lvert\alpha_r\rvert^2\Omega^2t^2\kappa^2} \right)^k\\
		&\left( \frac{2ix_0}{1+ 4\lvert\alpha_r\rvert^2\Omega^2t^2\kappa^2}  \right)^{n-2k} \lvert n\rangle,
	\end{align*}  
	where $x_0 = \arcsin\left[(n_d -n_c)/(n_d + n_c)\right]$. The behavior of the sum over $k$ dictates how strong the correlations would be. For instance, where the relative photon number difference $\rvert n_d - n_c\lvert$ is large and comparable to total photon number $n_c + n_d$, the magnitude of $x_0$ is large with a maximum value $\pi/2$. However, $\lvert\alpha_{r}\rvert^2$ is large compared to $\pi/2$. Thus, the term with the power $n- 2k$ is small compared to the term with power $k$, and is treated as a perturbation. Hence, maximum correlation is obtained in the limit $\lvert n_d - n_c\rvert \ll n_c + n_d$ for which $\lvert x_0\rvert \approx 0$, and $k$ takes on a value of $n/2$. The unnormalized state of the atoms conditioned on the detection of photons~(\ref{eq:conditionalstate}) then becomes
	\begin{align}
		\label{eq:approximatestate}
		\lvert\tilde{\psi}_{n_c,n_d}\rangle =& \frac{e^{-\frac{\lvert\alpha_l\rvert^2 + \lvert\alpha_r\rvert^2}{2}}}{\sqrt{n_c!}\,\sqrt{n_d!}}\left(\frac{n_c}{n_c + n_d}\right)^{\frac{n_c}{2}} \left(\frac{n_d}{n_c + n_d}\right)^{\frac{n_d}{2}}\nonumber\\
		&\times\frac{(\lvert \alpha_r\rvert^2 + \lvert\alpha_l\rvert^2)^{\frac{n_c + n_d}{2}}}{\sqrt{1 + 4\lvert\alpha_{r}\rvert^2 \Omega^2 t^2\kappa^2}} \sum_{n = 0,2,4,\cdots}\frac{\sqrt{n!}}{\left(\dfrac{n}{2}\right)!}\nonumber\\
		&\times i^n\left(\frac{2 \lvert\alpha_{r}\rvert^2\Omega^2t^2\kappa^2}{1 + 4 \lvert\alpha_{r}\rvert^2\Omega^2t^2\kappa^2}\right)^{\frac{n}{2}}\lvert n\rangle.
	\end{align}
	A global phase factor has been neglected in writing~(\ref{eq:approximatestate}). The final state obtained is not a spin coherent state. More so, the state $\lvert\psi_{n_c,n_d}\rangle$ accepts only even integer numbers --- the final state conditioned on photon detection is a superposition of even number states only. Such a superposition of the same parity states is characteristic of a squeezed state. The term $\dfrac{2 \lvert\alpha_{r}\rvert^2\Omega^2t^2\kappa^2}{1 + 4 \lvert\alpha_{r}\rvert^2\Omega^2t^2\kappa^2}$ determines the depth of the squeezing. For short atom-light interaction times, $\dfrac{2 \lvert\alpha_{r}\rvert^2\Omega^2t^2\kappa^2}{1 + 4 \lvert\alpha_{r}\rvert^2\Omega^2t^2\kappa^2}$ is small (less than unity), and  the probability amplitude of the  $\lvert 0\rangle$ is largest and goes to zero for larger $n$. Similarly, for long interaction times, $\dfrac{2 \lvert\alpha_{r}\rvert^2\Omega^2t^2\kappa^2}{1 + 4 \lvert\alpha_{r}\rvert^2\Omega^2t^2\kappa^2}$ is appreciable (greater than unity), and  the state at $n=N$ has the largest probability amplitude, while the amplitude of $n=0$  is smallest. In this case, it is obvious that the Holstein-Primakoff approximation of the atomic state is invalid. Hence, the validity of the Holstein-Primakoff approximation performed on the atomic state implies that $\dfrac{2 \lvert\alpha_{r}\rvert^2\Omega^2t^2\kappa^2}{1 + 4 \lvert\alpha_{r}\rvert^2\Omega^2t^2\kappa^2} < 0.5$ and is the regime where the atomic state is squeezed. The weak form of the condition is $ gt < 2(\lvert\alpha_{r}\rvert\sqrt{N})^{-1}$, which we put in the form that is amenable to the results obtained later below as $\Omega t < (\lvert\alpha_{r}\rvert\kappa)^{-1}$ where $\kappa$ is defined in (\ref{eq:strn:cv04}).
	
	The probability $P(n_c,n_d) = \langle \psi_{n_c,n_d}\lvert\psi_{n_c,n_d}\rangle$ of getting $n_c$, $n_d$ photons  in the measurement is obtained by summing over the \emph{n}th atom state, which converges for $ gt < 2(\lvert\alpha_{r}\rvert\sqrt{N})^{-1}$. Hence, the probability $P(n_c,n_d)$ becomes
	\begin{align}
		\label{eq:papc09}
		P(n_c,n_d) &= \frac{e^{-\lvert\alpha_l\rvert^2 -\lvert\alpha_r\rvert^2}}{n_c!\,n_d!}\left(\frac{n_c}{n_c + n_d}\right)^{n_c} \left(\frac{n_d}{n_c + n_d}\right)^{n_d}\nonumber\\
		& \times (\lvert \alpha_r\rvert^2 + \lvert\alpha_l\rvert^2)^{n_c + n_d} \frac{1}{\sqrt{1 + 8\lvert\alpha_{r}\rvert^2\Omega^2t^2\kappa^2}}.
	\end{align} 
	Consequently the normalized state~(\ref{eq:approximatestate}) of the atoms conditioned on photon detection is
	\begin{equation}
		\label{eq:normalizedstate}
		\begin{split}
			\lvert\psi_{n_c,n_d}\rangle & = \frac{\left(1 + 8\lvert\alpha_{r}\rvert^2\Omega^2t^2\kappa^2\right)^{1/4}}{\sqrt{1 + 4\lvert\alpha_{r}\rvert^2 \Omega^2 t^2\kappa^2}} \sum_{n=0 }^{N_\mathrm{max}}\frac{\sqrt{(2n)!}}{n!}\\
			&\times \left(\frac{-2 \lvert\alpha_{r}\rvert^2\Omega^2t^2\kappa^2}{1 + 4 \lvert\alpha_{r}\rvert^2\Omega^2t^2\kappa^2}\right)^{n}\lvert2 n\rangle,
		\end{split}
	\end{equation} 
	where $N_\mathrm{max} = N/2$ for $N$ even, and $N_\mathrm{max} = (N-1)/2$  for $N$ odd.
		
	Previously,  Refs.~\citep{ilo-okeke2014,ilo-okeke2016} reported that Faraday imaging is minimally-destructive for atom-light interaction strengths that are sufficiently small compared to the square-root of the number of atoms in the condensate, $gt\ll 1/\sqrt{N}$ which provides an upper bound. Here, the strength of the measurement and the point of observation (the photon number $n_c$ and $n_d$) would determine the required coupling strength $gt$. For a particular measurement  to be minimally destructive, it is the total number of atoms $N$ and the average photon number $\lvert\alpha_r\rvert^2$ that determine the strength of the atom-light coupling, $gt < 2/(|\alpha_r|\sqrt{N})$. Although the inequality is obtained in the undepleted pump approximation and for optimal performance,  it provides the condition of applicability of an ac Stark shift to be minimally destructive in a conditional measurement.
	
	%%%==========================================================
	%%% Purification of flawed qubits
	%%%==========================================================
	\subsection{Hybrid Variance\label{sec:sec:squeezing}}
	%%%========================================================== 
	The normalised state immediately after the detection of $n_c$ and $n_d$ photons is thus 
	\begin{align}
		\label{eq:pap03}
		\lvert\psi_{n_c,n_d}\rangle & = \frac{ e^{-\frac{\lvert\alpha_l\rvert^2 +\lvert \alpha_r\rvert^2}{2}}}{\sqrt{n_c!}\sqrt{n_d!}}\left(\frac{\alpha_r}{\sqrt{2}}\right)^{n_c+n_d}\frac{1}{\sqrt{P(n_c,n_d)}}\nonumber\\
		&\times \sum_{n_l=0}^\infty e^{in_l\phi}
		I(n_c,n_d,n_l) \nonumber\\
		&\times e^{ i\Omega t\left[\frac{\hat{X}^2 + \hat{P}^2}{2} + \sqrt{2} \kappa (n_c+n_d-2n_l) \hat{X}\right] }\lvert\psi_0\rangle, 
	\end{align}
	where $P(n_c,n_d)$ is the probability (\ref{eq:papc09}), and $I(n_c, n_d, n_l)$ is the Fourier coefficient (\ref{eq:strn:cv09a}). The normalized conditional expectation value $\bar{B}$ of any of the atom operator $\hat{B} =\hat{X},\,\hat{P}$ subject to having counted $n_c$ and $n_d$ photons at the detector is calculated using  $\lvert\psi_{n_c,n_d}\rangle$ as
	\begin{align}
		\label{eq:pap06}
		\bar{B} &= \sqrt{\frac{2}{N}} \langle\psi_{n_c,n_d}\lvert \hat{B}\rvert\psi_{n_c,n_d}\rangle.
	\end{align}		
	Similarly, the normalized conditional variance of the atom operators is found using (\ref{eq:pap06}) as
	\begin{equation}
		\label{eq:papc23}
		%			\begin{split}
			\Delta\bar{B} =2\left[ \langle\psi_{n_c,n_d} \lvert \hat{B}^2\lvert\psi_{n_c,n_d}\rangle 
			-\left( \langle\psi_{n_c,n_d}\lvert \hat{B}\lvert\psi_{n_c,n_d}\rangle\right)^2\right].
			%			\end{split}
	\end{equation} 
	In the text, equations (\ref{eq:pap06}) and (\ref{eq:papc23}) will be referred to as the hybrid expectation values and hybrid variance, respectively. The variances are normalized such that for $g=0$, $\Delta\bar{B}$ gives unity, $\Delta\bar{B} =1$.
	
	%%%==========================================================
	%%% Purification of flawed qubits
	%%%==========================================================
	\subsection{Approximate  Variance \label{sec:sec:ApproximateExpression}}
	%%%========================================================== 
	The details of calculating the mean values and the variances are given in Appendix~\ref{sec:operators} which we summarize here. Using (\ref{eq:strn:cv03}) in (\ref{eq:pap06}), we first calculate the approximate expression for the hybrid mean values of  $\hat{X}$, and $\hat{P}$ (\ref{eq:strn:cv02}). The expectation values are evaluated at $n_c, \,n_d = (\lvert\alpha_r\rvert^2 + \lvert\alpha_l\rvert^2)/2$, and $\alpha_l =\alpha_r$. Because $\lvert \psi_{n_c,n_d}\rangle $ selects only state with even parity, we find for an \emph{n}th state $\lvert n \rangle$, $\langle n\rvert\hat{\mathcal{O}}\lvert\psi_{n_c,n_d}\rangle =0$ for \emph{n} even where $\hat{\mathcal{O}} = b^\dagger_e,\,b_e$, while $\langle n\rvert \psi_{n_c,n_d}\rangle \neq 0$ for \emph{n} even, and vice versa. Hence, the expectation value of operator $\hat{\mathcal{O}}$ is easily calculated by summing over $n$, $\langle\psi_{n_c,n_d}\rvert \hat{\mathcal{O}}\lvert\psi_{n_c,n_d}\rangle = \sum_n\langle \psi_{n_c,n_d}\lvert n\rangle\langle n\lvert \hat{\mathcal{O}}\rvert\psi_{n_c,n_d}\rangle$. It is immediately deduced from (\ref{eq:pap06}) that the expectation values of $\hat{X}$ and $\hat{P}$ are
	\begin{equation}
		\label{eq:papc22}
		\langle\psi_{n_c,n_d}\lvert \hat{X}\rvert\psi_{n_c,n_d}\rangle = \langle\psi_{n_c,n_d}\lvert \hat{P}\rvert\psi_{n_c,n_d}\rangle = 0, 
	\end{equation}
	for all time. For the operators $\hat{\mathcal{O}} = b^\dagger_e b_e, \,(b^\dagger_e)^2,\,\mathrm{and} \, b^2_e$ needed in the calculation of variances, $\langle n\lvert \hat{\mathcal{O}} \rvert\psi_{n_c,n_d}\rangle \neq 0 $ for \emph{n} even, and zero otherwise. Hence, their expectation values are evaluated as $\langle\psi_{n_c,n_d}\lvert\hat{\mathcal{O}}\rvert\psi_{n_c,n_d}\rangle = \sum_n\langle\psi_{n_c,n_d}\lvert n\rangle \langle n\lvert\hat{\mathcal{O}}\rvert\psi_{n_c,n_d}\rangle$. Using these, the normalized approximate variances of $\hat{X}$ and $\hat{P}$ calculated using (\ref{eq:papc23}) evaluate to
	\begin{align}
		\label{eq:papc24}
		\Delta \bar{X} & = \frac{1}{1 + 8\lvert\alpha_{r}\rvert^2\Omega^2t^2\kappa^2} ,\\
		\label{eq:papc25}
		\Delta \bar{P} & =  1 + 8\lvert\alpha_{r}\rvert^2\Omega^2t^2\kappa^2.
	\end{align}
	Because of the normalization implemented in (\ref{eq:papc23}), the approximate variances~(\ref{eq:papc24}) and~(\ref{eq:papc25}) are such that for $g=0$, they give $1$.  Hence a division by $2$ recovers the expected values of $1/2$. We note that the variances of the quadrature $X$ and $P$ [(\ref{eq:papc24}) and (\ref{eq:papc25}), respectively] do not depend on the tunneling frequency $\Omega$. This is  because the parameter $\kappa$ defined in (\ref{eq:strn:cv04}) is proportional to the inverse of the tunneling frequency $\Omega$, $\kappa = g\sqrt{N}/(2\Omega)$. However, to be able to make comparisons with the results of Ref.~\cite{ilo-okeke2021}, we will keep their present form.

	%==================================================		
	\subsection{Numerical Evaluation \label{sec:sec:ExactSolution}}
	%==================================================
	We now numerically solve the Schr\"odinger equation~(\ref{eq:papc29}) to compare the results that we have obtained in the previous section to exact results. At the end of the evolution, the photons are interfered by the beamsplitter according to the transformation (\ref{eq:OperatorTransformation}). The detection of $n_c$ and $n_d$ photons following immediately their interference collapses the wave function of light and is described by the projection operator $\mathcal{P} = \lvert n_c,n_d \rangle \langle n_c,n_d\rvert$. The details of the numerical method for solving the Schr\"odinger equation is given in Appendix~\ref{sec:SchrdingerEquation}.  The state $\lvert \Psi_{n_c,n_d}\rangle$ of the atoms, as given in Eq. (\ref{eq:strn:me10}), is conditioned upon the detection of light and is used to calculate the normalized averages of atomic spin variables $J_{x,y}$: 
	\begin{align}
		\label{eq:ExactMean}
		\bar{J}_i &= \frac{2}{N}\langle \Psi_{n_c,n_d}\lvert J_i\rvert\Psi_{n_c,n_d}\rangle,\\
		\label{eq:ExactVariance}
		\Delta\bar{J}_i &= \frac{4}{N}\left(\langle \Psi_{n_c,n_d}\lvert J_i^2\rvert \Psi_{n_c,n_d}\rangle  - \frac{N^2}{4}\bar{J}_i^2\right),
	\end{align}
	for $i = x,\, y$. These are used later in Secs.~\ref{sec:sec:ResultPerfectPolarization},~\ref{sec:sec:ExpectationValuesVsExact} and~\ref{sec:sec:VariancesOfAtomicOperators}, and are  the exact solution for the mean values and variances, respectively.

	%%%%%===========================================
	\subsection{Exact Solution Versus The Hybrid Approximation Results \label{sec:sec:ResultPerfectPolarization}}
	%%%%%===========================================
	%%===========================================
	%%Figure 2
	%%===========================================
	\begin{figure}[t]
		\includegraphics[width=\columnwidth]{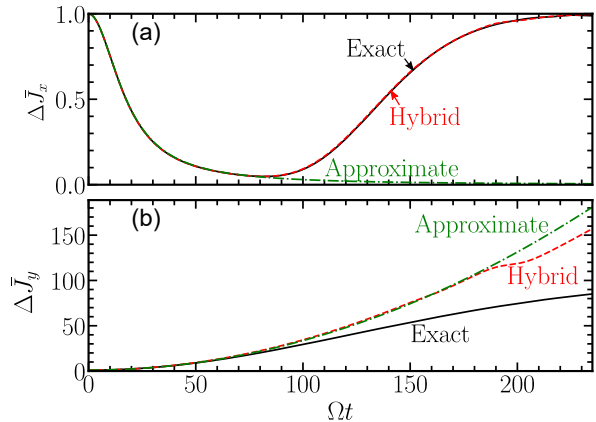}
		\caption{The normalized conditional variance of $J_x$, $J_y$. In (a), the solid line is exact solution (\ref{eq:papc29}) calculated as described in the text. The dashed line is hybrid variance obtained using (\ref{eq:strn:cv03}) in (\ref{eq:papc23}), while the dashed-dotted line is the approximate variance (\ref{eq:papc24}). 	In (b), the solid line is the exact variance (\ref{eq:papc29}) calculated as described in the main text. The dashed line is the hybrid variance obtained using (\ref{eq:strn:cv03}) in (\ref{eq:papc23}), while the dashed-dotted line is the approximate variance (\ref{eq:papc25}).	The parameters for plots are $\alpha_l = \alpha_r = \sqrt{20}$, $N = 200$, $\Omega = \pi/4$, and $g = 0.1\Omega/N$.}
		\label{fig13}
	\end{figure}	
	%%===========================================
	
	The results for the variance calculations are presented in Fig.~\ref{fig13}. The variance of  $\hat{X}$ in Eq. (\ref{eq:papc23}) is evaluated using the hybrid approximation, decreases below unity with an increase in time before returning to unity after longer time scales as shown in Fig.~\ref{fig13}(a). These are in perfect agreement with the exact variance  $J_x$ in Eq. (\ref{eq:ExactVariance}). On the other hand, the hybrid variance $\Delta\bar{P}$ is never below unity and agrees with the exact variance $\Delta\bar{J}_y$ for the short time interval $\Omega t < (\lvert\alpha\rvert\kappa)^{-1}$ as shown in Fig.~\ref{fig13}(b). To better understand this behavior we turn to approximate expression for the variance of $\hat{X}$ and $\hat{P}$ given in (\ref{eq:papc24}) and (\ref{eq:papc25}), respectively. In Fig.~\ref{fig13}(b), it is seen that in the short time regime, the approximate variance $\Delta \bar{P}$ (Eq. (\ref{eq:papc25})) and  hybrid variance (\ref{eq:papc23}) agree with the exact solution $\Delta\bar{J}_y$ (Eq. (\ref{eq:ExactVariance})). The variance of $J_y$ increases quadratically with time at a rate given by $\lvert\alpha_{r}\rvert^2\Omega^2\kappa^2$ as given by the approximate variance $\Delta\bar{P}$ (Eq. (\ref{eq:papc25})).

	Similarly, for short times, the approximate variance $\Delta\bar{X}$ (Eq. (\ref{eq:papc24})) starts out decreasing quadratically with time below unity roughly at a rate $\lvert\alpha_{r}\rvert^2\Omega^2\kappa^2$ in good agreement with the hybrid variance (\ref{eq:papc23}), and the exact solution $\Delta\bar{J}_x$ (Eq. (\ref{eq:ExactVariance})), as shown in Fig.~\ref{fig13}(a). The decrease of the variance below unity shows that the state is squeezed, as the error in the mean value is below the classically expected, which in normalized units is unity. This confirms the existence of correlations in the condensate as discussed in Sec.~\ref{sec:sec: probden}, see (\ref{eq:normalizedstate}). However, at longer times $\Omega t \geq (|\alpha_r|\kappa)^{-1}$, the approximate variance (Eq. (\ref{eq:papc24})) decreases inversely with the square of time roughly at a rate $(\lvert\alpha_{r}\rvert^2\Omega^2\kappa^2)^{-1}$, and goes to zero as $1/t^2$. In the same time regime,  the hybrid variance $\Delta \bar{X}$ and exact solution $\Delta \bar{J}_x$ reach their minimum value and then increase towards unity. It is not  surprising that the approximate variance $\Delta \bar{X}$ breaks down. As already discussed in Sec.~\ref{sec:sec: probden}, the validity of approximate results hold for  $\Omega t < (\lvert \alpha_{r}\rvert \kappa)^{-1}$. It is in this neighborhood that the exact variance $\Delta\bar{J}_x$ reaches its minimum (best) squeezing value and the approximate variances $\Delta\bar{X}$ (Eq. (\ref{eq:papc24}))  and $\Delta\bar{P}$ (Eq. (\ref{eq:papc25})) break down. 
	
	Note that in the region of importance where there is a squeezing effect, where the exact solution $\Delta\bar{J}_x$ is decreasing with time, the approximate variance $\Delta\bar{X}$ (Eq. (\ref{eq:papc24})) does well predicting its behavior. It starts failing in the neighborhood of the minimum squeezing as predicted by the exact solution $\Delta\bar{J}_x$ (Eq. (\ref{eq:ExactVariance})), where the gains made in error reduction are beginning to be lost. Hence one can always predict the onset of losing squeezing gained from the measurement for times not fulfilling the condition $g t <2(\sqrt{N} \lvert\alpha_r\rvert)^{-1}$. Similarly, the hybrid variance~(\ref{eq:papc23}) and approximate expression ~(\ref{eq:papc25}) of $\Delta\bar{P}$ start to deviate from the exact solution $\Delta\bar{J}_y$ (Eq. (\ref{eq:ExactVariance})) in the same time neighbourhood, $g t \sim 2 (\sqrt{N} \lvert\alpha_r\rvert)^{-1}$ as shown in Fig.~\ref{fig13}(b).   Surprisingly, the hybrid variance~(\ref{eq:papc23}) does well in predicting the exact solution~(\ref{eq:ExactVariance})  even at long times, $g t > 2(\sqrt{N} \lvert\alpha_r\rvert)^{-1}$.
	
	Ordinarily one would expect the system to exhibit oscillations for atoms tunneling in a double-well trap. The lack of oscillations in the solutions of $J_x$ and $J_y$  is due to the initial state~(\ref{eq:strn:cv03}) used in their calculation. For this state, the contribution of $\hat{X}^2 + \hat{P}^2\propto\hat{b}^\dagger_e\hat{b}_e$, which is responsible for oscillation, is zero, see Appendix.~\ref{sec:disentangle}. Thus its exponential operator does not affect the vacuum state and does not contribute to the averages of spin operators.

	%%%%==========================================================
	%%%% Purification of flawed qubits
	%%%%==========================================================
	\subsection{Husimi \emph{Q}-distribution \label{sec:sec:results}}
	%%%%==========================================================
	%%===========================================
	%%Figure 3
	%%===========================================
	\begin{figure}[t]
		\includegraphics[width=\columnwidth]{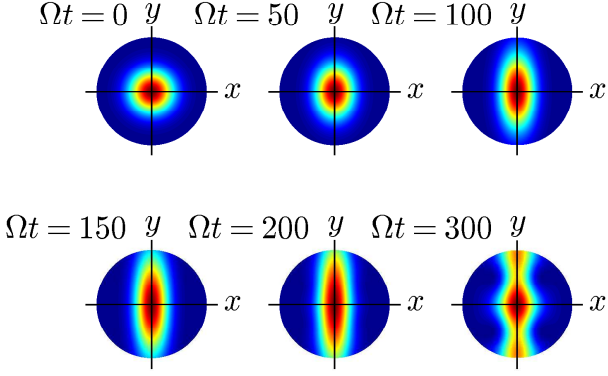}
		\caption{The conditional Husimi \emph{Q}-function. The parameters for the plots are $\alpha_l = \alpha_r = \sqrt{12}$, $N = 1000$, $\Omega = \pi/4$, and $g=0.1\Omega/N$. The time $\Omega t$ for each plot is as shown in the figure.} 
		\label{fig12}
	\end{figure}
	%%===========================================
	As shown in Sec.\ref{sec:sec: probden}, measurements induce correlations in the condensates. They manifest as a reduction in the variance below the standard quantum limit for the atomic spin variable coupled to light, as shown in Sec.\ref{sec:sec:ResultPerfectPolarization}. Additionally, correlations affect the width of the distribution representing the quantum state. To visualize the effect of correlation among the atoms on the state $\lvert \psi_{n_c,n_d}\rangle$ (\ref{eq:pap03}) we plot the Husimi \emph{Q}-function defined as 
	\begin{equation}
		\label{eq:pap17}
		Q_{n_c,n_d}(\nu,t) = \frac{\lvert\langle\nu\lvert\psi_{n_c,n_d} \rangle\rvert^2}{\pi},
	\end{equation}
	where $\lvert\nu \rangle= e^{-\tfrac{|\nu|^2}{2}} e^{\nu b^\dagger_e}\lvert 0\rangle$, is the coherent state of atoms within the Holstein-Primakoff approximation~\cite{byrnes2021} [see also (\ref{eq:papc26})], and $\nu = \lvert\nu\rvert e^{i\varphi}$. The \emph{Q}-function is numerically calculated and is shown in Fig.~\ref{fig12}. The distribution starts out at $\Omega t = 0$ as a circle on \emph{x}--\emph{y} plane. This is the typical behavior of a coherent state with a normalized width of unity. As time progress, the circle becomes squeezed along the \emph{x}--axis. However, at time $\Omega t = 200$, the width of the distribution starts showing a subtle increase in the width along \emph{x}--axis. The squeezing along \emph{x}--axis is lost at $\Omega t = 300$. To better understand these effects, we explicitly calculate the conditional Husimi \emph{Q}-function  
	\begin{align}
		\label{eq:pap18}
		Q_{n_c,n_d}(\nu,t)& = \frac{e^{-|\alpha_l|^2 - |\alpha_r|^2 -|\nu|^2}}{\pi\,n_c!\, n_d!}\frac{\left(|\alpha_r|^2 + |\alpha_l|^2\right)^{n_c+n_d}}{P(n_c,n_d)}\nonumber\\
		&\times\left(\frac{ n_c}{n_c + n_d} 	\right)^{n_c}\left(\frac{n_d}{n_c + n_d} \right)^{n_d}\frac{1}{1 + 4\sigma\Omega^2 t^2 \kappa^2}\nonumber\\
		&\times e^{-\frac{\Omega^2t^2\kappa^2(n_c+n_d -2a)^2}{1 + 	4\sigma\Omega^2t^2\kappa^2}} e^{-\frac{\sigma(x_0 - \phi)^2}{1 + 4\sigma \Omega^2t^2\kappa^2}} \nonumber\\
		&\times e^{-\frac{4\sigma\Omega t\kappa \lvert \nu\rvert(x_0 	- \phi)\cos\varphi}{1 + 4\sigma \Omega^2t^2\kappa^2}} e^{-\frac{1 + 8\sigma\Omega^2t^2\kappa^2\cos^2\varphi}{1 + 4\sigma\Omega^2 t^2\kappa^2}\lvert\nu\rvert^2}\nonumber\\
		&\times e^{\frac{2\Omega t\kappa(n_c + n_d - 	2a)\vert\nu\rvert\sin\varphi}{1 + 4\sigma\Omega^2t^2\kappa^2}}.
	\end{align} 
	
	Equation (\ref{eq:pap18}) shows that there are contributions to the \emph{Q} function due to the meter (laser light) and the detection in addition to the effect that we intend to observe that is proportional to $\nu^2$ in the exponent. The exponents with term $(x_0 -\phi)$ contain  contributions from the relative phase of the measurement arising from the relative phase $\phi = \phi_l - \phi_r$ of the light beams in the arms of the interferometer, and the measured (detected) phase $x_0$ (see (\ref{eq:strn:cv18})) and $\sigma$ is defined in~(\ref{eq:strn:cv28}). The exponents with the term $n_c + n_d -2a = (n_c + n_d)(\rvert\alpha_r\rvert^2 -\rvert\alpha_l\rvert^2)/(\rvert\alpha_r\rvert^2 + \rvert\alpha_l\rvert^2)$ represent the contributions from the observed total photon number $n_c + n_d$. These contributions can be made to vanish hence limiting the effect of measurement on the state of the atoms~\cite{ilo-okeke2021}. For instance, at balanced detection, the amplitude and phase of light beams in the two arms of the interferometer are same, $\phi_l = \phi_r$ and $\lvert\alpha_r\rvert = \lvert\alpha_l\rvert$. Consequently, the relative phase $\phi$ and the fractional total population $n_c+n_d -2a$ identically vanish, $\phi =0 = n_c+n_d -2a$. Additionally, where the measured relative photon number difference is small compared to the total photon number, ($
	\rvert n_d -  n_c\lvert \ll n_c + n_d $), $x_0 \approx 0$. With these conditions and upon substituting (\ref{eq:papc09}) in (\ref{eq:pap18}), it simplifies to  
	\begin{align}
		\label{eq:pap19}
		Q_{n_c,n_d}(\nu,t) = \frac{1}{\pi }\frac{\sqrt{1 + 			8\lvert\alpha_{r}\rvert^2\Omega^2 t^2\kappa^2}}{1 + 4\lvert\alpha_{r}\rvert^2\Omega^2t^2\kappa^2}
		\,e^{-\frac{1 + 8\lvert\alpha_{r}\rvert^2\Omega^2t^2\kappa^2\cos^2\varphi}{1 + 		4\lvert\alpha_{r}\rvert^2\Omega^2t^2\kappa^2}\nu^2},
	\end{align}  
	where the effect of measurement on \emph{Q} function is contained in its width $\Delta\nu$.
	
	It is then evident that the \emph{Q} function has a peak at $\nu =0$ with width $\Delta\nu$ given by 
	\begin{equation}
		\label{eq:pap20}
		\Delta\nu^2 = \frac{1 + 4\lvert\alpha_{r}\rvert^2\Omega^2t^2\kappa^2}{1 + 	8\lvert\alpha_{r}\rvert^2\Omega^2t^2\kappa^2\cos^2\varphi}.
	\end{equation}
	As expected, the measurement has no effect on the width of the \emph{Q} function for $t=0$. In this case, the width is simply unity, the same to that of the coherent state $\lvert \psi_0\rangle$ (Eq. (\ref{eq:strn:cv03})), which is equal in all directions. This is shown in Fig.~\ref{fig12} at $\Omega t = 0$. The limit $\Delta\nu =1$ is the so-called standard quantum limit. For $t >0$, the phase-angle $\varphi$ controls the direction on \emph{x}--\emph{y} plane. For instance $\varphi = \pi/2,\,3\pi/2$ corresponds to the positive and negative \emph{y}--axis respectively. Similarly, $\varphi=0,\,\pi$ corresponds to the positive and negative \emph{x}--axis. It then becomes clear that the width along the \emph{y}--axis (both negative and positive) increases quadratically with time, roughly at the rate $\lvert\alpha_{r}\rvert^2\Omega^2\kappa^2$. 
	
	Similarly, the width along the \emph{x}--axis (both positive and negative) decreases quadratically with time from unity for small times $0<\Omega t \ll 1$ at a rate $\lvert\alpha_{r}\rvert^2\Omega^2\kappa^2$. These cause the width shrink along the \emph{x}--axis while extending along the \emph{y}--axis. Hence the quantum state $\lvert \psi_{n_c,n_d}\rangle$ is said to be squeezed along \emph{x}--axis. This is shown in Fig.~\ref{fig12} at $\Omega = 50, 100, 150$. Additionally, (\ref{eq:pap20}) predicts finite squeezing as the width along \emph{x}--axis tends to $\Delta\nu^2 =1/2$ as $\Omega t$ becomes very large, which is the case as seen from Fig.~\ref{fig12} at $\Omega t = 150$. A careful study of the plot at $\Omega t = 200$ shows a subtle increase in the width of the distribution along the \emph{x}--axis. These imply that the gains made are gradually being eroded with an increase in time. This is seen clearly in Fig.~\ref{fig12} at $\Omega t = 300$. That (\ref{eq:pap20}) breaks down---predicting finite squeezing with infinite time $\Omega t \gg (|\alpha_r|\kappa)^{-1}$---is not surprising as the expression is only valid for $g t < 2 (\sqrt{N}\lvert\alpha_r\rvert)^{-1}$, which is the region where the width of the distribution is squeezed as discussed in Secs.~\ref{sec:sec: probden} and~\ref{sec:sec:ResultPerfectPolarization}.

	%%%========================================================== 
	\section{Imperfect initial spin polarization\label{sec:comparison}}
	%%%=========================================================
	In many experimental situations, there may be a finite but small atomic population in the excited state of the double-well trap. In these situations, the initial state is not perfectly polarized. Such a situation could lead to results qualitatively different from perfectly polarized states, as discussed in Sec.~\ref{sec:results}. In this section, using a coherent state as the initial state for imperfect polarization, we establish the mechanism for the existence of quantum squeezing correlations at short interaction time and their loss at long interaction time. We then calculate the averages of atomic spin variables.

	%=====================================
	\subsection{Emergence And Disappearance Of Squeezing Correlations\label{sec:sec:ExcitedStateNotZero}}
	%=====================================
	Consider the case where there is a finite but small fractional population of atoms in the excited state of the double-well trap, $0<|\alpha|^2\ll1$. In this limit, the initial state~(\ref{eq:AtomCoherentState}) reduces to  \cite{byrnes2021}
	\begin{equation}
		\label{eq:papc26}
		\lvert \psi_0 \rangle = 	e^{-N\frac{\lvert\alpha\rvert^2}{2}}\sum_{m=0}^{N} \frac{(\sqrt{N}\alpha)^m}{\sqrt{m!}}\lvert m\rangle ,
	\end{equation} 
	for the undepleted pump approximation.  
	
	The passage of light through the condensate and subsequent detection of photons as shown in Fig.~\ref{fig11} gives the unnormalized state of the atoms~(\ref{eq:conditionalstate}) where the initial state $\lvert\psi_0\rangle$ is (\ref{eq:papc26}). The action of the exponential operator $e^{i\Omega t\left[\frac{\hat{X}^2 + \hat{P}^2}{2} + \sqrt{2}\kappa(n_c+n_d-2n_1)\hat{X}\right]}$ on the state $\lvert\psi_0\rangle$ is to first impart a phase proportional to the tunneling frequency $\Omega$, then displace it by an amount $-i\Omega t \kappa (n_c + n_d - 2n_1) $, see~(\ref{eq:dis03}),
	\begin{align}
		\label{eq:displacedstate}
		\lvert\psi_{n_c,n_d}\rangle& \propto \sum_{n_l=0}^{\infty} e^{in_l\phi} I(n_c,n_d,n_l) e^{-i\Omega t \kappa (n_c + n_d - 2n_l)\sqrt{N}\alpha}\nonumber\\
		&\times e^{-\frac{N\lvert\alpha\rvert^2 + \Omega^2 t^2 \kappa^2 (n_c + n_d - 2n_l)^2 }{2}} \nonumber\\
		& \times\sum_{m=0}^{N} \frac{\left(-i\Omega t \kappa (n_c + n_d - 2n_l) + \sqrt{N}\alpha e^{i\Omega t}\right)^m}{\sqrt{m!}}\lvert m\rangle.
	\end{align}
	Additionally, the exponential term in the second line proportional to $N\lvert\alpha\rvert^2$ broadens the width of the initial state. A spin-coherent-like state emerges if $N\rvert\alpha\lvert^2 $ is dominant while a correlated state emerges where $\lvert\alpha_{r}\rvert^2 \Omega^2\kappa^2t^2$ is the dominant term, as we show below.
	
	The unnormalized state of the atoms obtained by summing over $n_l$ under the conditions that the measurement be balanced $\alpha_{l} = \alpha_{r}$ and the relative photon number difference in a measurement be small in comparison with the total number of photons, $\rvert n_d - n_c\lvert \ll n_d + n_d$ is
	\begin{align}
		\label{eq:unnormalizedstate}
		\lvert \tilde{\psi}_{n_c,n_d}\rangle &= \frac{e^{-\frac{\lvert\alpha_l\rvert^2 + \lvert\alpha_r\rvert^2}{2}}}{\sqrt{n_c!}\,\sqrt{n_d!}}\left(\frac{n_c}{n_c + n_d}\right)^{\frac{n_c}{2}} \left(\frac{n_d}{n_c + n_d}\right)^{\frac{n_d}{2}}\nonumber\\
		&\times\frac{(\lvert \alpha_r\rvert^2 + \lvert\alpha_l\rvert^2)^{\frac{n_c + n_d}{2}}}{\sqrt{1 + 4\lvert\alpha_{r}\rvert^2 \Omega^2 t^2\kappa^2}} e^{-\frac{N\lvert\alpha\rvert^2}{2}} e^{-\frac{2\lvert\alpha_{r}\rvert^2\Omega^2t^2\kappa^2 N\alpha^2}{(1 + 4\lvert\alpha_{r}\rvert^2\Omega^2t^2\kappa^2)}}\\
		&\times \sum_{n=0}^{N} \sum_{k=0}^{n/2}\frac{\sqrt{n!}}{k!(n-2k)!} \left(\frac{-2 \lvert\alpha_{r}\rvert^2\Omega^2t^2\kappa^2}{1 + 4 \lvert\alpha_{r}\rvert^2\Omega^2t^2\kappa^2}\right)^k\nonumber\\
		&\times\left(\sqrt{N}\alpha\left[e^{i\Omega t}-\frac{4 \lvert\alpha_{r}\rvert^2\Omega^2t^2\kappa^2}{1 + 4 \lvert\alpha_{r}\rvert^2\Omega^2t^2\kappa^2}\right]\right)^{n-2k} \lvert n\rangle\nonumber,
	\end{align}
	where we have used the completeness relation $\sum_n \lvert n\rangle\langle n\rvert = \hat{\mathbbm{1}}$ in evaluating the sum. Equation (\ref{eq:unnormalizedstate}) presents two interesting limits for the state $\lvert\psi_{n_c,n_d}\rangle$. In one of the limits  where $\kappa = 0$ (or $gt = 0$), there is no interaction between the atoms and light, the unnormalized state~(\ref{eq:unnormalizedstate}) is simply the product of the collapsed state $\lvert n_c,n_d\rangle$ and the initial state (\ref{eq:papc26}). Similarly, for $\sqrt{N}\alpha$ being small but large in comparison with atom-light coupling strength $\lvert\sigma\rvert\Omega t\kappa$, then the terms with exponent $n-2k$  is the dominant term ($|e^{i\Omega t}|\sim 1$). Hence, the terms with exponent $k$ is treated as a perturbation, and $k$ takes on values $0,1,\cdots,n/2$ for $n$ even and $0,1,\cdots, (n-1)/2 $ for $n$ odd. Nevertheless, whether $n$ is even or odd only a few terms $k=0$ and possibly $k=1$ are sufficient to describe the amplitude of the state $\lvert n\rangle$ in this limit. Hence no correlations exist among the atoms, and the physics of the system can be described classically.
	
	In the other limit where $\sqrt{N}\alpha$ is small compared to the atom light coupling strength $\lvert\alpha_r\rvert\Omega t\kappa$, then the term with the exponent $n-2k$ is treated as a perturbation. In this limit, there exist correlations among the atoms. As a special case where $\alpha =0$, $k$ must be $n/2$ and one recovers (\ref{eq:approximatestate}), see also (\ref{eq:normalizedstate}). For $\sqrt{N}\lvert\alpha\rvert\ll 1$ and $gt <2(\lvert\alpha_{r}\rvert\sqrt{N})^{-1}$, and considering leading and the next term in $k $ ($k = n/2, n/2-1$), the normalized state conditioned on the detection of photons becomes
	\begin{equation}
		\label{eq:approximatenormalizedstate}
		\begin{split}
			\lvert \psi_{n_c,n_d}\rangle & \approx \frac{\left(1 + 8\lvert\alpha_{r}\rvert^2\Omega^2t^2\kappa^2\right)^{1/4}}{\sqrt{1 + 4\lvert\alpha_{r}\rvert^2 \Omega^2 t^2\kappa^2}} \frac{1}{\sqrt{P_\epsilon}} \sum_{n=0}^{N_\mathrm{max}}  \Bigg[\frac{\sqrt{(2n)!}}{n!}\\
			&  \times\left(\frac{-2 \lvert\alpha_{r}\rvert^2\Omega^2t^2\kappa^2}{1 + 4 \lvert\alpha_{r}\rvert^2\Omega^2t^2\kappa^2}\right)^{n} +  N\alpha^2\\
			& \times\frac{\sqrt{(2n)!}}{2(n-1)!}\left(\frac{-2 \lvert\alpha_{r}\rvert^2\Omega^2t^2\kappa^2}{1 + 4 \lvert\alpha_{r}\rvert^2\Omega^2t^2\kappa^2}\right)^{n-1} \\
			&\times \left(e^{i\Omega t}-\frac{4 \lvert\alpha_{r}\rvert^2\Omega^2t^2\kappa^2}{1 + 4 \lvert\alpha_{r}\rvert^2\Omega^2t^2\kappa^2}\right)^2\Bigg]\lvert 2n\rangle,
		\end{split}
	\end{equation}
	where $N_\mathrm{max} = N/2$ for $N$ even, $N_\mathrm{max} = (N-1)/2$  for $N$ odd and  the corrections to the probability is
	\begin{equation}
		\label{eq:approxamp}
		P_\epsilon \approx 1 - N\rvert\alpha\lvert^2\frac{4\lvert\alpha_{r}\rvert^2\Omega^2t^2\kappa^2}{1+8\lvert\alpha_{r}\rvert^2\Omega^2t^2\kappa^2}\cos\left(2\Omega t + 2\varepsilon\right),
	\end{equation}
	and $\varepsilon$ is the argument of $\alpha$. Again, we see that the case $\alpha =0$ (\ref{eq:normalizedstate}) is a special case.  It is then evident that the state (\ref{eq:approximatenormalizedstate}) is a superposition of even number states thereby affirming the existence of correlations in the condensate. The state (\ref{eq:approximatenormalizedstate}) is a squeezed state of an atomic oscillator.
	
	At small atom-light interaction strengths, the effect that is observed depends on  which term has the greater value---the initial population $N\lvert\alpha\rvert^2$ or the atom-light interaction strength $gt$. For instance, starting with the conditions that the atom-light interaction strength is smaller than the initial population $N\lvert\alpha\rvert^2$, there are no correlations. However, because the parameter $gt$ depends on time, these conditions are dynamic. Eventually, correlations emerge in the regime where atom-light interaction strengths become greater than the initial population of atoms. Hence, having very small population of atoms  in the excited state $N\lvert\alpha\rvert^2\ll 1$ ensures that correlations always exist at small interaction times.
	
	There is a loss of squeezing correlations at long interaction times, as already noted in Sec.~\ref{sec:sec: probden}. This is because at long atom interaction times, the Fourier term (\ref{eq:strn:cv09a}) presents other peaks besides $x_0$, see for instance $x_0'$ in (\ref{eq:strn:cv32}). These other peaks that were exponentially small at short interaction times and thus negligible in the evaluation of atomic wave function are now comparable to $x_0$. Physically, this implies that $x_0$ is no longer the only dominant detectable amplitude---other amplitudes are now possible. The interference (superposition) of these amplitudes scrambles the squeezing correlations which leads to their loss at a long interaction times. This effect was also observed in Fig.~2(b) of Ref.~\cite{ilo-okeke2021}. However, the superposition of amplitudes gives rise to other forms of non-Gaussian correlations, such as the Schr{\"o}dinger cat state~\cite{ilo-okeke2021} that emerge at long interaction times. Hence, the loss of squeezing is a precursor to the emergence of other non-Gaussian correlations. 
	
	%		It manifests as loss  of squeezing as would be observed in the variances calculated below.

	We thus see that for systems which are not perfectly polarized, when the population executing the  oscillation $N\lvert\alpha\rvert^2$ exceeds the atom-light coupling strength $gt$, no correlations will be observed. The photon statistics dominates its physics and the system behaves classically. Correlations begin to emerge where the population $\sqrt{N}\alpha$ is comparable to interaction strength $gt$. Consequently, correlations dominate the physics of the system only in the regime where $\sqrt{N}\lvert\alpha\rvert\ll 1$.  Since the coherent state is a state of an oscillator, it then implies that quantum effects such as correlations may be observed in oscillators using QND measurement, if the number of oscillators are small, $N\lvert\alpha\rvert^2\ll 1$.  
	
	%		\begin{align}
		%			\label{eq: }
		%			\lvert\psi_{n_c,n_d}\rangle & = \frac{ e^{-\frac{\lvert\alpha_l\rvert^2 +\lvert \alpha_r\rvert^2}{2}}}{\sqrt{n_c!}\sqrt{n_d!}}\left(\frac{\alpha_r}{\sqrt{2}}\right)^{n_c+n_d}\frac{1}{\sqrt{P(n_c,n_d)}}\sum_{n_1}e^{in_1\phi}
		%			\nonumber\\
		%			&I(n_c,n_d,n_1) e^{ i\Omega t\left[\frac{\hat{X}^2 + \hat{P}^2}{2} + \sqrt{2} \kappa (n_c+n_d-2n_1) \hat{X}\right] }\lvert\psi_0\rangle. 
		%		\end{align}
	
	%=====================================
	\subsection{The Conditional Expectation Values \label{sec:sec:ExpectationValuesVsExact}}
	%=====================================
	%%===========================================
	%%Figure 14
	%%===========================================
	\begin{figure}[t]
		\includegraphics[width=\columnwidth]{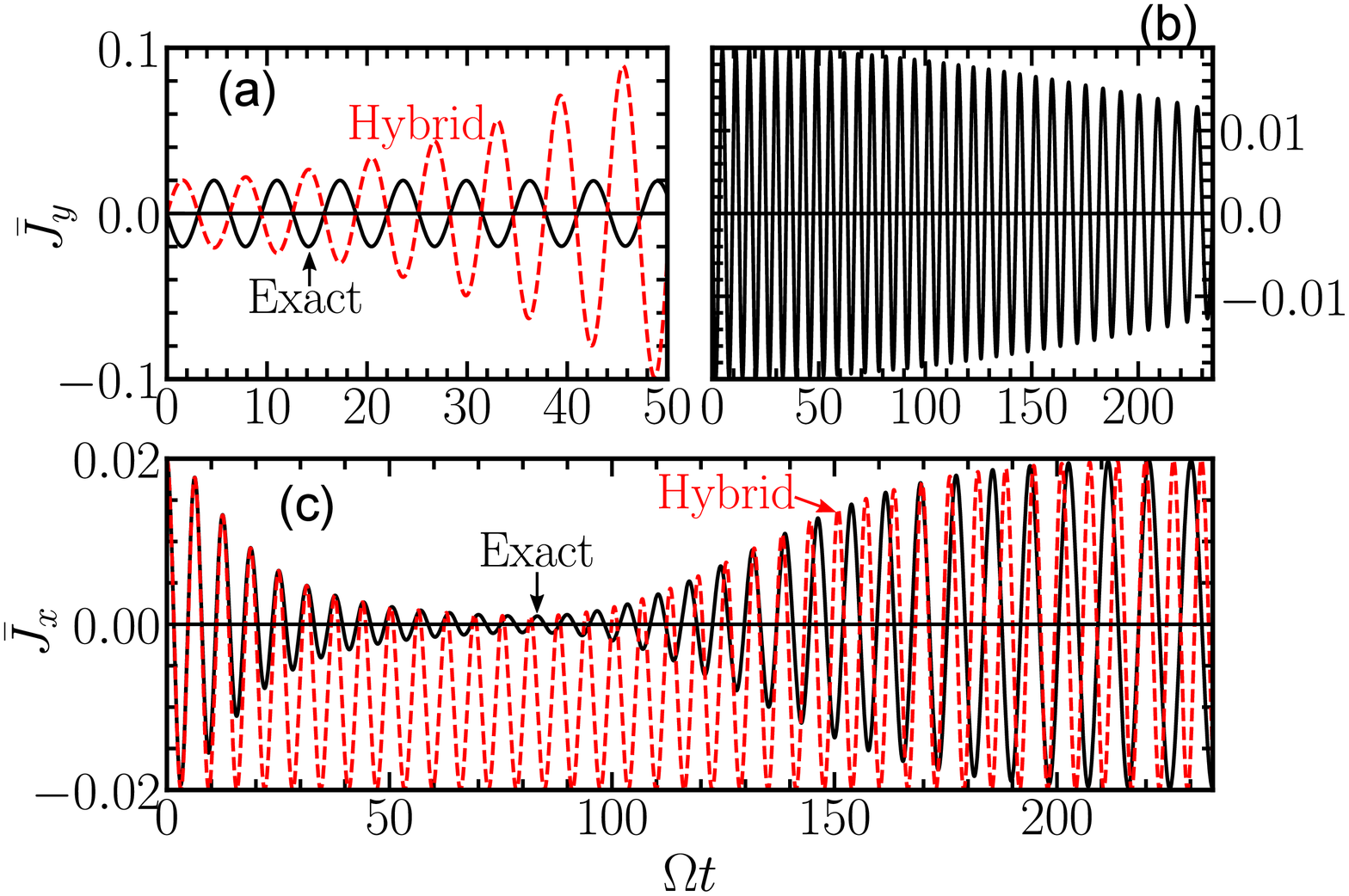}
		\caption{The normalized conditional mean values of $J_x$, and $J_y$. In (a), the exact mean (solid line) is calculated using (\ref{eq:ExactMean}) while the  hybrid mean (dashed line) is calculated using (\ref{eq:papc26}) in (\ref{eq:pap06}). (b) shows the behavior of the exact mean value  $\bar{J}_y$ for the entire evolution.  In (c), the exact mean (solid line) is calculated using (\ref{eq:ExactMean}) while the hybrid mean (dashed line) is calculated using (\ref{eq:papc26}) in (\ref{eq:pap06}). The parameters for plots are $\alpha_l = \alpha_r = \sqrt{20}$, $N = 200$, $\Omega = \pi/4$ and $g = 0.1\Omega/N$, $\beta=\sqrt{9999}\times 10^{-2} $ and $\lvert\alpha\rvert = 0.01$.}
		\label{fig14}
	\end{figure}
	%%===========================================
	
	Using the initial state~(\ref{eq:papc26}), we numerically calculate the averages for $\hat{X}$ and $\hat{P}$ and compare the results with the exact solution. It is seen from Fig.~\ref{fig14} that the exact mean values of $J_x$ and $J_y$ oscillate about the hybrid mean, the solid line at $\bar{J}_x =0$ and $\bar{J}_y=0$, respectively, when $\alpha=0$. The measurement provides a modulation envelope while the tunneling is responsible for the oscillations. As shown in Fig.~\ref{fig14}(c), the exact solution tends to zero in the region where modulation by measurement is strong. In addition, the exact solution oscillates about zero and only approaches it linearly, as shown in Figs.~\ref{fig14}(a)~and~(b). 
	
	Some of this behavior can be deduced from the hybrid mean values. For instance, the measurement modulates the oscillations of the exact solution $\bar{J}_x$ and $\bar{J}_y$.  At short interaction times, $\Omega t  < (\lvert\alpha_{r} \rvert\kappa)^{-1}$, we see from Fig.~\ref{fig14}(c) that the oscillations in the hybrid calculation are in phase with the exact solution. Additionally, the amplitude of oscillations of the hybrid means is not symmetric about zero.  This is because the modulation by the measurement causes the hybrid mean value to approach zero for the amplitude of oscillation greater than zero, while the amplitude below zero is unaffected. The approximate mean obtained in the short time limit, $\Omega t  < (\lvert\alpha_{r} \rvert\kappa)^{-1}$, while treating the oscillations as perturbation $\sqrt{N}\lvert\alpha \rvert\ll 1$, equally shows this behavior
	\begin{align}
		\label{eq:papc27}
		\bar{X}_\alpha \approx 	&\frac{\lvert\alpha\rvert}{(1+4\lvert\alpha_{r}\rvert^2\Omega^2\kappa^2t^2)^2} \Big[-8\lvert\alpha_{r}\rvert^2\Omega^2\kappa^2t^2 \cos\phi_\alpha \nonumber\\
		& + 2(1+ 	4\lvert\alpha_{r}\rvert^2\Omega^2\kappa^2t^2)\cos(\Omega t + \phi_\alpha)  \Big],
		%	\langle \hat{X}\rangle \approx \frac{\lvert\alpha'\rvert}{\sqrt{2}} \Big[-8\sigma\Omega^2\kappa^2t^2 \cos\phi_\alpha + 2\cos(\Omega t + \phi_\alpha)\Big]
	\end{align}
	where $\phi_\alpha$ is the phase of the fractional population in the excited state. It is zero in all cases considered in this work. For $\cos(\Omega t + \phi_\alpha)$ positive, notice that there are cancellations between the first and second term that leads to the reduction in the amplitude of the hybrid mean. However, when $\cos(\Omega t + \phi_\alpha)$ is negative, the terms add. This effect brings about the one-sided modulation effect exhibited by the expectation value of the hybrid mean in Fig.~\ref{fig14}(c).
	
	On the other hand, we see that in Fig.~\ref{fig14}(a) the amplitude of the hybrid mean value increases with every oscillation while being modulated by the measurement, with modulation strength proportional to $\lvert\alpha_{r}\rvert^2\Omega^2\kappa^2 t^2$. This effect is opposite to that exhibited by the amplitudes of the exact mean value. The oscillations decrease very slowly, as shown in Fig.~\ref{fig14}(b). The poor prediction of the behavior of the  exact solution by hybrid mean value is on display in the approximate mean value, which under the same conditions as (\ref{eq:papc27}), gives 
	\begin{align}
		\label{eq:papc28}
		\bar{P}_\alpha &\approx 	\frac{\lvert\alpha\rvert\left(1 + 8\lvert\alpha_{r}\rvert^2\Omega^2\kappa^2t^2\right)}{\left(1+ 4\lvert\alpha_{r}\rvert^2\Omega^2\kappa^2t^2\right)^2}\Big[-8\lvert\alpha_{r}\rvert^2\Omega^2\kappa^2t^2 \sin\phi_\alpha\nonumber\\
		& + 2(1 + 	4\lvert\alpha_{r}\rvert^2\Omega^2\kappa^2t^2)\sin(\Omega t+\phi_\alpha )\Big].
	\end{align}  
	Since $\phi_\alpha=0$, it is immediately evident that $\langle\hat{P}\rangle$ is modulated by the measurement and increases quadratically with time in the short time regime. Additionally, the oscillations in the hybrid mean value is $\pi$ out of phase with the exact solution as shown in Fig.~\ref{fig14}(a). 
	
	The hybrid approximation does not capture all the features of the exact mean values of atom operators. It is notably worse for predictions of the hybrid mean value about $\bar{J}_y$. The failure of the hybrid approximation to describe the exact mean values in the presence of oscillations stems from approximating the operators of the ground state by a c-number. Such an approximation is good when the ground state of the oscillator contributes no more than random fluctuations. In this case, the excited state alone sufficiently describes the physics of the system. However, for a small but finite population of atoms in the excited state $0<\sqrt{N}\lvert\alpha\rvert \ll 1$, the ground state contribution can no longer be approximated by a c-number. Consequently, the excited state alone is insufficient in describing the mean values of the system. Including corrections to hybrid approximation would not improve the results since the next order correction is proportional to order three in terms of the operators, $\propto b^{\dagger2}_eb_e,\propto b^{\dagger}_eb_e^2$ and vice versa, hence will give rise to terms of order $N^{3/2}\lvert\alpha\rvert^3$ which is less than $\sqrt{N}\lvert\alpha\rvert$ for $\sqrt{N}\lvert\alpha\rvert <1$.

	%%===========================================
	%%Figure 15
	%%===========================================
	\begin{figure}[t]
		\includegraphics[width=\columnwidth]{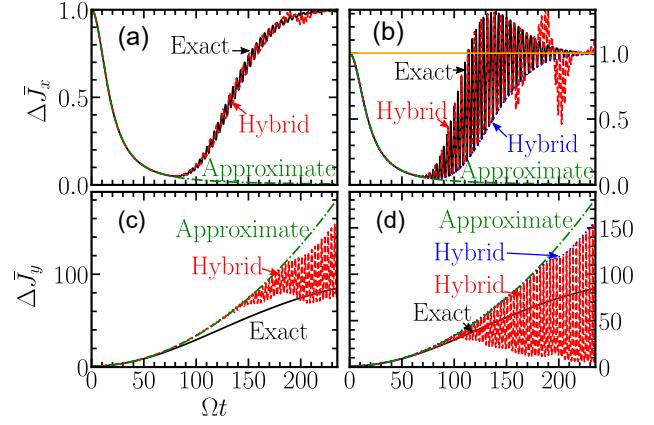}
		\caption{The normalized conditional variance of $J_x$ and $J_y$. (a) The exact solution $J_x$ (solid line) is calculated from (\ref{eq:ExactVariance}) as described in the main text. The dashed line is the hybrid variance calculated using (\ref{eq:papc26}) in (\ref{eq:papc23}), while the dashed-dotted line is the approximate variance (\ref{eq:papc24}). (b) The exact variance (solid line)is calculated in same way as in~(a) and similarly for the hybrid variance (dashed line). The approximate variance (dashed-dotted line) is from (\ref{eq:papc24}) while the hybrid variance (dotted line) calculated using (\ref{eq:strn:cv03}) in (\ref{eq:papc23}). The straight line at $\Delta\bar{J}_x = 1$ is a guide for the eye. In (c), the exact variance (solid line) is calculated from (\ref{eq:ExactVariance}) as described in the main text. The hybrid variance (dashed line) is calculated using (\ref{eq:papc26}) in (\ref{eq:papc23}), while the  dashed-dotted is the approximate variance (\ref{eq:papc25}). In (d), the exact variance of $J_y$ (solid line) is calculated in same way as in~(c) and similarly for the hybrid variance (dashed line). The dashed-dotted line is approximate variance  (\ref{eq:papc25}) while the hybrid variance (dotted line) is calculated using (\ref{eq:strn:cv03}) in (\ref{eq:papc23}). The parameters for plots are $\alpha_l = \alpha_r = \sqrt{20}$, $N = 200$, $\Omega = \pi/4$ and $g = 0.1\Omega/N$. For the first column, $\beta=\sqrt{9999}\times 10^{-2} $ and $\lvert\alpha\rvert = 0.01$, while for the second column $\beta=\sqrt{0.999}$ and $\alpha = \sqrt{0.001}$.} 
		\label{fig15}
	\end{figure}
	%=====================================================================
	
	%=====================================
	\subsection{Conditional Variances Of Atomic Operators  \label{sec:sec:VariancesOfAtomicOperators}}
	%=====================================
	We now compare the normalized variance $\Delta\bar{J}_y$ and its hybrid approximation in Fig.~\ref{fig15}(c). In the short time regime, $\Omega t < (\lvert\alpha_r\rvert\kappa)^{-1}$, there is a very good agreement between the exact solution (solid line) and the hybrid variance (dashed line). For a better understanding of the results, we turn to the approximate expression (\ref{eq:papc25}) valid for short times and obtained by assuming that the oscillator amplitude is small in comparison to unity, $\sqrt{N}\lvert\alpha\rvert\ll1$. At short times, there is no dependence of the variances on the oscillations, and the approximate variance (\ref{eq:papc25}) predicts that the exact solution~(\ref{eq:ExactVariance}) and hybrid variance~(\ref{eq:papc23}) increase quadratically with time. This is in remarkable agreement with the results of the hybrid approximation and the exact solution, as shown in  Fig~\ref{fig15}(c). For long times, $\Omega t> (\lvert \alpha_{r}\rvert \kappa)^{-1}$, however, the hybrid variance and the approximate variance are qualitatively close but do not agree precisely with the exact variance. In particular, there are oscillations present in the hybrid variance that are not present in either approximate variance or the exact solution~(\ref{eq:ExactVariance}). In the regime where the amplitude of oscillation is large, $\sqrt{N}\lvert\alpha\rvert <1$, as shown in Fig.~\ref{fig15}(d), there is still agreement between the hybrid variance and the exact solution in the short time regime, $\Omega t < (\lvert \alpha_{r}\rvert\kappa)^{-1}$. But still, there are no oscillations in the exact solution for a long time, which are very prominent in the hybrid variance.  
	
	The exact variance $\Delta\bar{J}_x$ (solid line), and hybrid approximation are compared in Fig.~\ref{fig15}(a). There is very good agreement between the exact solution~(\ref{eq:ExactVariance}) and hybrid variance (\ref{eq:papc23}) in the squeezing regime, $\Omega t < (\lvert \alpha_{r}\rvert\kappa)^{-1}$. These results are understood by turning to the approximate variance (\ref{eq:papc24}) obtained for short times, $gt < 2(\sqrt{N}\lvert\alpha_r\rvert)^{-1}$, while treating the oscillation amplitude as a perturbation, $\sqrt{N}\lvert\alpha\rvert \ll 1$. There are no oscillations in this regime. The approximate variance predicts that the variance of $J_x$ decreases quadratically with time which is in remarkable agreement with the exact solution and hybrid variance in the squeezing regime $gt < 2 (\sqrt{N}\lvert\alpha_r\rvert)^{-1}$. It starts failing at the onset of losing squeezing where $gt <2 (\sqrt{N}\lvert\alpha_r\rvert)^{-1}$ is no longer satisfied. Beyond this regime, squeezing is being lost, and there are oscillations in the exact solution and the hybrid variance. However, the oscillations in the hybrid variance occur at every other cycle of the exact solution. Hence, the exact solution oscillates at approximately twice the frequency of the hybrid variance. Notice that the amplitude of these oscillations is small and hence is well below unity.
	
	However, for an appreciable number of atoms in the excited state, the oscillator amplitude will be appreciable though small, $\sqrt{N}\lvert\alpha\rvert <1$. This leads to squeezing and anti-squeezing effects occurring in the regime $\Omega t > (\lvert \alpha_{r}\rvert\kappa)^{-1}$ as shown in Fig.~\ref{fig15}(b). The minima of the variances shown in Fig.~\ref{fig15}(a) or Fig.~\ref{fig15}(b) are points where there are no oscillations. For instance, as shown in Fig.~\ref{fig15}(b) the minima of oscillations in exact solution and hybrid variance correspond to the values of the hybrid variance calculated using (\ref{eq:strn:cv03}) in (\ref{eq:papc23}). It suggests treating oscillations as a perturbation to the non-oscillatory state $\alpha=0$. Hence, systems with large oscillations are strongly perturbed. In general, we posit that the hybrid approximation does better in predicting the variances of operators than it did in predicting their mean values.
	
	The success of the hybrid approximation in describing variance within the short time regime is understood in the following way. The conditional variance of the system is one-half, irrespective of the value of $\alpha $. In the presence of atom-light interactions, the correction term happens to be the contribution from the measurement. Higher order corrections consisting of the product of $\lvert\alpha_r\rvert^2\Omega^2\kappa^2 t^2 N\lvert\alpha\rvert^2$ are negligible. Hence the zeroth order terms from the oscillator and first-order correction provided by the measurement are sufficient to describe the variances of system operators, especially in the regime where there is squeezing. In the long interaction time regime, terms in Fourier coefficient $I(n_c,n_d,n_l)$~(\ref{eq:strn:cv09a}) which were exponentially small or not present in the short term regime are no longer small and become comparable to the term used in the short interaction time regime, see for instance~(\ref{eq:strn:cv32}). These terms contributed to corrections that improved the overall behavior of the hybrid approximation, especially for the spin variable coupled to light.

	%=====================================
	\section{Composition of Variance in Squeezing Regime\label{sec:discussion}}
	%=====================================
	From the foregoing analysis of the variances with or without oscillations, it is apparent that the variance of any of the system's operator $\hat{C}$ in the squeezing regime, $gt <(\sqrt{N}\lvert\alpha_r\rvert)^{-1}$, can, in general, be written as the sum of the variance of the operators $(\Delta C)_0^2$ in the absence of measurement and the variance of the meter 
	\begin{equation}
		\label{eq:papc30}
		(\Delta C)^2 = (\Delta C)_0^2 + (\Delta \mathrm{meter})^2.
	\end{equation}
	This is exemplified in the approximate expressions given in (\ref{eq:papc24}) and (\ref{eq:papc25}). For a collection of a large number $N$ of particles, the variance of system operators will be near their classical limit, $(\Delta C)_0^2\sim N^{-1}$. Squeezing is realized for the system operator that couples to the meter by a suitable choice of the meter parameters such that there are cancellations between the variance of that system operator $(\Delta C)_0^2$ and the variance of the meter $(\Delta \mathrm{meter})^2$. Squeezing is lost in the regime where the measurement strength is comparable to the inverse product of the average photon number and the square root of the number of particles, $gt \sim 2(\sqrt{N}\lvert\alpha_r\rvert)^{-1}$. Beyond the squeezing regime, one can still observe squeezing even though squeezing is being lost as discussed in Sec.~\ref{sec:sec:squeezing}. For an initial state of the system prepared with a finite but small population in the excited state, there will be oscillations beyond the squeezing regime as discussed in  Sec.~\ref{sec:sec:ExpectationValuesVsExact}. If the population of atoms in the excited state is finite but small $N\alpha \ll 1$, it is still possible to observe squeezing beyond the squeezing regime as the amplitude of the oscillations will be small. However, for an excited state that is appreciably populated $N\alpha < 1$, there will be oscillations that may be present in the squeezing regime. Because of the large amplitude of oscillations in the regime where squeezing is being lost, there will be times when the variance $(\Delta C)^2$ is greater than the initial variance $(\Delta C)_0^2$, thus being anti-squeezed as discussed in Sec~\ref{sec:sec:ExcitedStateNotZero}. To observe squeezing in this regime may require filters or timing the measurement to do away with the anti-squeezing.
	 
	%%%==========================================================
	\section{Experimental Implementation  \label{sec:implementation}}
	%%%==========================================================
	We previously detailed possible implementations of our model in experiments in Ref.~\cite{ilo-okeke2021}. Here, we briefly describe an experimental realization of our model. A BEC in a radio-frequency double-well potential~\cite{harte2018,barker2020} realizes the atom oscillator described in Fig.~\ref{fig11}. Squeezing the quantum state of the atom oscillator can then be done by working in the frequency domain. Because atoms in the minima of the double well experience different trapping frequencies,  they modulate an off-resonant light field~\cite{jammi2018} passing through them at different frequencies, thus realizing single-site addressing. A balanced polarimeter detects the signal at each frequency~\cite{jammi2018}. We did not consider effects due to photodetection efficiency that limits the efficacy of resolving the number of photons collected by the detectors~\cite{barnett1998,lee2004}. It affects the quantum state of photons by decreasing the linear superposition of the average photon numbers $ (\lvert \alpha_{l}\rvert^2 + \lvert\alpha_{r}\rvert^2)$ by an amount proportional to detection efficiency $\eta_E$, $(\lvert \alpha_{l}\rvert^2 + \lvert\alpha_{r}\rvert^2)\rightarrow\eta_E (\lvert \alpha_{l}\rvert^2 + \lvert\alpha_{r}\rvert^2)$. By doing so, it masks the number of photons at the readout so that the actual values $n_c,\, n_d$ are never known, thereby introducing randomness in the readout. As a result, each QND measurement readout is random and thus is affected by the single-particle resolution of the detectors. Here, we do not need to know the actual values of the number of photons at the readout, and our results retain their form in the presence of inefficient photodetection. As such, squeezing would still be observed in the presence of imperfect photodetection. However, decoherence from the spontaneous emission of the atoms would modify the results of this work~\cite{ilo-okeke2021}.
	
% 	We did not consider effects due to photo-detection efficiency that limits efficacy in resolving the number of photons collected by the detectors~\cite{barnett1998,lee2004}. In quantum state preparation using QND, the detection of photons happens later, after preparing the atomic state by light. Hence, any effect from imperfect photodetection would affect the quantum state of photons by causing each photon number state to evolve at a different rate, thereby scrambling the quantum state of light in a manner reminiscent of photon loss. Since the photon and oscillator states are entangled, imperfect single-photon resolution manifests on an atomic variable or its quantum state obtained by averaging all possible photon number outcomes as a decay in the measured variable or amplitude of the quantum state. On the contrary, the effects of imperfect photodetection do not manifest where there is no averaging as described here since those effects cancel out in the normalized quantum state of the atom oscillator.
	
	%%%==========================================================
	\section{Summary and conclusions  \label{sec:summary}}
	%%%==========================================================
	We analyzed the squeezing effect in a double-well potential using the hybrid approximation, where the spin variables of the atoms are treated within the Holstein-Primakoff approximation, but the photonic degrees of freedom are calculated exactly. We considered an initial state where all the atoms are in the ground state of the double-well trap. The main result of this paper is~(\ref{eq:normalizedstate}), which shows that correlations develop because of the measurement. The correlations manifest in the state of the atoms as a superposition of even number states. We calculated simple expressions for the variances of atom operators, (\ref{eq:papc24}) and (\ref{eq:papc25}). We showed that in the squeezing regime, the variance is the sum of the variances of the system and the meter (\ref{eq:papc30}). The hybrid approximation agrees very well with the exact solution obtained by solving the Schr\"odinger equation in the short interaction time regime.  We found the short interaction time regime to be where the atom-light interaction strength $gt$ is comparable to the inverse of the product of the average photon number $\lvert\alpha_{r}\rvert$ and the square-root of the total atom number $N$ in the trap, $gt \sim 2 (\lvert \alpha_{r}\rvert \sqrt{N})^{-1}$. It marks the onset of losing squeezing, and the approximate results failed beyond the short interaction time regime.

	Additionally, we calculated the Husimi \emph{Q}-distribution and highlighted various contributions from the light and its detection to the \emph{Q}-distribution. We showed that these contributions vanished for a balanced detection and a readout performed at zero photon number difference. The Husimi \emph{Q}-distribution showed that the quantum state of the atom condensates is squeezed. We obtained a simple expression (\ref{eq:pap20}) that explains the variations in the width of the distribution in phase space as the BEC state goes from a coherent state to being squeezed.
	
	We further checked the validity of the hybrid approximation in the presence of a small but finite population of atoms in the excited state of the double-well trap. We showed that correlations emerge depending on the atom-light interaction strength being greater than the population of atoms in the excited state~(\ref{eq:unnormalizedstate}). The implication is that quantum features such as correlations are not observed in oscillators using QND measurement unless the number of oscillators is small. We calculated the averages of the spin variables. The approximate variances agreed very well with the exact solution in the squeezing regime, $gt <2 (\lvert \alpha_r\rvert \sqrt{N})^{-1}$.  However, the hybrid mean values performed poorly in predicting the exact solution. The hybrid mean value for the spin variable coupled to light agreed well with the modulation of the exact solution. For the spin variable not coupled to light, the hybrid mean was $\pi/2$ out of phase with the exact solution. Additionally, the hybrid mean predicted increases in the modulation with time, while the exact solution decreased linearly over time.   
	 
	 The emergence of superposition of amplitudes in the quantum state of the condensate, beyond the short time regime $gt > 2 (\lvert \alpha_r\rvert \sqrt{N})^{-1}$, led to the loss of squeezing correlations, and heralds the emergence of non-Gaussian correlations~\cite{pezze2018,ilo-okeke2021}. Additionally, it provides corrections to the variance of the spin variable coupled to light. Hence, the hybrid variance well reproduces the results of the exact solution. However, the superposition of amplitudes did not improve the variance of the spin variables not coupled to light. Hence, the hybrid approximation failed to replicate the exact variance for the spin variable not coupled to light.  Nonetheless, the onset of losing squeezing predicted as $gt \sim 2/(\sqrt{N}|\alpha_r|)$ in the hybrid approximation worked very well for the exact solution.

	\begin{acknowledgments}
		This work is supported by the National Natural Science Foundation of China (62071301); NYU-ECNU Institute of Physics at NYU Shanghai; the Joint Physics Research Institute Challenge Grant; the Science and Technology Commission of Shanghai Municipality (19XD1423000,22ZR1444600); the NYU Shanghai Boost Fund; the China Foreign Experts Program (G2021013002L); the NYU Shanghai Major-Grants Seed Fund; Tamkeen under the NYU Abu Dhabi Research Institute grant CG008.
	\end{acknowledgments}

	\appendix
	%%==========================================================
	\section{Solving The Schr\"odinger Equation\label{sec:SchrdingerEquation}}
	%%==========================================================
	To find the evolution of the the state vector $\Psi(t)$, we use the ansatz
	\begin{equation}
		\label{eq:strn:me02}
		\lvert \Psi(t)\rangle = \sum_{k}\psi_{k}(t)\lvert k;\alpha_{k,l}(t),\alpha_{k,r}(t)\rangle,
	\end{equation}
	where for each atom number state $\lvert k\rangle = \lvert k,N-k\rangle$ defined in the $J_x$ basis, there is a coherent state $\lvert\alpha_{k,s} \rangle= e^{\tfrac{-|\alpha_{k,s}|^2}{2}}e^{\alpha_{k,s} \hat{a}^\dagger_s}\lvert 0\rangle$, $s=l,r$, and $\lvert k;\alpha_{k,l},\alpha_{k,r}\rangle = \lvert k\rangle\otimes \lvert\alpha_{k,r}\rangle\otimes \lvert\alpha_{k,l}\rangle$. Substituting the ansatz, (\ref{eq:strn:me02}) into (\ref{eq:papc29}) produces two states of light $\lvert \alpha_{k,s}\rangle$ and $\hat{a}^\dagger \lvert\alpha_{k;l,s}\rangle$ that are not necessarily orthogonal. Decomposing  the state $\hat{a}^\dagger \lvert\alpha_{k,s}\rangle $ into its orthogonal form  $\hat{a}^\dagger \lvert\alpha_{k,s}\rangle = \lvert \alpha_{\perp,k,s}\rangle + \alpha^*_{k,s}\lvert\alpha_{k,s}\rangle$ where $\langle \alpha_{\perp,k,s} \lvert\alpha_{k,s}\rangle =0$ gives the equations for the evolution of $\alpha_{k,s}$ as 
	\begin{align}
		\label{eq:strn:me03}
		\frac{d\alpha_{k,l}}{dt }&= - \frac{ig(2k - N)}{2}\alpha_{k,l},\nonumber\\ 
		\frac{d\alpha_{k,r}}{dt }&=  \frac{ig(2k - N)}{2}\alpha_{k,r}, 
	\end{align}
	whose solutions are 
	\begin{align}
		\label{eq:papc04}
		\alpha_{k,l}(t) & = \alpha_{l} e^{- i(2k-N)\tfrac{gt}{2}},\nonumber\\
		\alpha_{k,r}(t) & = \alpha_{r} e^{ i(2k-N)\tfrac{gt}{2}}.
	\end{align}
	Clearly we see that the effect of the atom-light interaction is to impact a phase on the photons. This phase carrying information can be extracted in an interference measurement. 
	
	Eliminating the $\alpha_{k,s}$ in the evolution of $\rho_{kk'}$ using~(\ref{eq:strn:me03}) and their solutions~(\ref{eq:papc04})~\cite{garcia-ripoll2005,hussain2014} give the following equation for the matrix elements $\rho_{kk'}$
	\begin{align}
		\label{eq:strn:me04}
		\frac{d\psi_k}{dt} & =   - i\Omega\frac{\sqrt{k(N-k+1)}}{2} \langle\alpha_{k}\lvert\alpha_{k-1}\rangle\psi_{k-1}\nonumber\\
		& - i\Omega\frac{\sqrt{(k+1)(N-k)}}{2} \langle\alpha_{k}\lvert\alpha_{k+1}\rangle\psi_{k+1},\nonumber\\
	\end{align}
	where $\langle \alpha_k\lvert\alpha_{k\pm1}\rangle$ is the overlap of light coherent state that interacted with $k$ atoms and light coherent state that interacted with $k \pm 1$, respectively,
	\begin{align}
		\label{eq:strn:me05}
		\langle\alpha_{k}\lvert\alpha_{k+1}\rangle &= e^{-(|\alpha_{l}|^2+|\alpha_{r}|^2)}e^{|\alpha_{l}|^2e^{-igt}}e^{|\alpha_{r}|^2e^{igt}},\nonumber \\
		\langle\alpha_{k}\lvert\alpha_{k-1}\rangle &= e^{-(|\alpha_{l}|^2+|\alpha_{r}|^2)}e^{|\alpha_{l}|^2e^{igt}} e^{|\alpha_{r}|^2e^{-igt}}.
	\end{align}
	Equation~(\ref{eq:strn:me04}) is a time dependent differential equation as determined by the phases (\ref{eq:strn:me05}). 
	
	At the beamsplitter, photons that left each well are recombined and sorted into bins $c$ and $d$ according to (\ref{eq:OperatorTransformation}). The state after recombination at the beamsplitter becomes
	\begin{align}
		\label{eq:strn:me08}
		\lvert\Psi\rangle_\mathrm{BS}(t) & = \sum_{k} \psi_{k}(t) 	e^{-\left(\frac{\lvert \alpha_{k,l}\rvert^2}{2}+\frac{\lvert \alpha_{k,r}\rvert^2}{2}\right)}e^{\frac{\alpha_{k,l} +i\alpha_{k,r}}{\sqrt{2}}\hat{a}^\dagger_c } \nonumber\\
		&\times e^{\frac{i\alpha_{k,l} 	+\alpha_{k,r}}{\sqrt{2}}\hat{a}^\dagger_d } \lvert k,0,0\rangle 
	\end{align}
	The probability of counting $n_c$ and $n_d$ photons in the bins $c$ and $d$, respectively, after recombination at the beamsplitter is obtained by the expectation of the projection operator $\mathcal{P}=\lvert n_c, n_d\rangle\langle n_c, n_d\rvert$ in the atoms subspace,
	\begin{align}
		\label{eq:strn:me09}
		P(n_c, n_d) & = \mathrm{Tr}[\lvert\Psi\rangle_\mathrm{BS}(t)\langle\Psi\rvert_\mathrm{BS}(t)\mathcal{P}],\nonumber\\
		& = \frac{1}{n_c!}\frac{1}{n_d!}\sum_{k}\psi_{k}^*\psi_{k}(t)e^{-\left(\lvert\alpha_{k,l}\rvert^2 + \lvert\alpha_{k,r}\rvert^2\right)}\nonumber\\ &\times\left(\frac{\lvert\alpha_{k,l} \rvert^2 +i\alpha_{k,r}\alpha^*_{k,l} - i\alpha_{k,l}\alpha^*_{k,r} + \lvert\alpha_{k,r}\rvert^2 }{2}\right)^{n_c}\nonumber\\
		&\times\left(\frac{\lvert\alpha_{k,l} \rvert^2 -i\alpha_{k,r}\alpha^*_{k,l} +i\alpha_{k,l}\alpha^*_{k,r} + \lvert\alpha_{k,r}\rvert^2 }{2}\right)^{n_d}.
	\end{align}
	
	Thus the state of the atoms given that $n_c$ and $n_d$ photons have been detected becomes
	\begin{align}
		\label{eq:strn:me10}
		\lvert \Psi_{n_c,n_d}\rangle &= \frac{1}{\sqrt{n_c!}}\frac{1}{\sqrt{n_d!}}\frac{1}{\sqrt{P(n_c,n_d)}}\sum_{k}\psi_{k}(t)\times\nonumber\\ &e^{-\left(\frac{\lvert\alpha_{k,l}\rvert^2}{2} +  \frac{\lvert\alpha_{k,r}\rvert^2}{2} \right)} \left(\frac{\alpha_{k,l} + i\alpha_{k,r} }{\sqrt{2}}\right)^{n_c}\nonumber\\
		&\times\left(\frac{i\alpha_{k,l} +\alpha_{k,r} }{\sqrt{2}}\right)^{n_d} \lvert k,N-k\rangle.
	\end{align}
	%
	
	%%%==========================================================
	\section{Disentangling Of Exponential Operators  \label{sec:disentangle}}
	%%%==========================================================
	Given the Hamiltonian (\ref{eq:strn:cv01}), we would like to write the exponential of the Hamiltonian as product of exponential of each term appearing in it. Such simplification allows for easy analysis of the exponential operator on a quantum state. We consider an operator of the form $U = e^{i\Omega t \left[\frac{\hat{X}^2 + \hat{P}^2}{2} -\sqrt{2}\kappa n\hat{X} \right]}$ appearing in Sec.~\ref{sec:cv} and Sec.~\ref{sec:results}. The operator may also be written in terms of $\hat{b}_e$ and $\hat{b}^\dagger_e$ 
	\begin{equation}
		\label{eq:dis01}
		U = e^{i\Omega t \left[\frac{\hat{X}^2 + \hat{P}^2}{2} 	-\sqrt{2}\kappa n\hat{X} \right]}= e^{i\Omega t \left[b^\dagger_eb_e  -\kappa n(\hat{b}^\dagger_e + \hat{b}_e) \right]},
	\end{equation}
	while we have ignored a constant factor of one-half that introduced an overall global phase in writing the second equality. The operators $\hat{b}^\dagger_e \hat{b}_e$, $\hat{b}^\dagger_e$ and $\hat{b}_e$ satisfy the following commutation relation
	\begin{align}
		\label{eq:dis02}
		[\hat{b}^\dagger_e \hat{b}_e,\hat{b}^\dagger_e]  & = 	\hat{b}^\dagger_e,\nonumber\\
		[\hat{b}^\dagger_e \hat{b}_e,\hat{b}_e] & = -\hat{b}_e,\\
		[\hat{b}_e,\hat{b}^\dagger_e ] &= 1.\nonumber
	\end{align} 
	Hence, any equivalent representation of $U$ would consist exponential operators made up of these operators and a constant. Thus, we seek to write $U $ in normally ordered form as 
	\begin{equation}
		\label{eq:dis03}
		U = e^{ip(\theta)\hat{b}^\dagger_e } 	e^{iq(\theta)\hat{b}^\dagger_e \hat{b}_e} e^{ir(\theta)\hat{b}_e }e^{is(\theta)} ,
	\end{equation}
	where $\theta = \Omega t$. Differentiating (\ref{eq:dis01}) with respect to $\theta$ gives 
	\begin{equation}
		\label{eq:dis04}
		\frac{d U}{d\theta} = i\left(b^\dagger_eb_e  -\kappa 	n(\hat{b}^\dagger_e + \hat{b}_e\right) U.
	\end{equation}
	Similarly, differentiating (\ref{eq:dis03}) with respect to $\theta$ gives
	\begin{align}
		\label{eq:dis05}
		\frac{d U}{d\theta} & =i \Bigg(s' + p'\hat{b}^\dagger_e + q' 	e^{ip\hat{b}^\dagger_e} b^\dagger_e\hat{b}_e e^{-ip\hat{b}^\dagger_e}  \nonumber\\
		& + r' e^{ip\hat{b}^\dagger_e} e^{iq\hat{b}^\dagger_e \hat{b}_e} 	\hat{b}_e e^{-iq\hat{b}^\dagger_e \hat{b}_e} e^{-ip\hat{b}^\dagger_e}  \Bigg) U
	\end{align}
	where the primes on $p$, $q$, $r$ and $s$ means differentiation with respect to $\theta$. 
	
	Applying Baker-Campbell-Hausdorff relation to (\ref{eq:dis05}), and equating coefficients of operators in (\ref{eq:dis04}) and (\ref{eq:dis05}), one obtains the following coupled differential equation for the functions,
	\begin{align}
		\label{eq:dis06}
		p' -iq'p & =-\kappa n,\nonumber\\
		r'e^{-iq} & = - \kappa n,\\
		q' & = 1, \nonumber\\
		s' - ir'pe^{-iq} & = 0. \nonumber
	\end{align}
	Solving the differential equations (\ref{eq:dis06}) with the initial condition $p(\theta=0) =r(\theta=0) = q(\theta=0) = s(\theta=0)=0$ gives
	\begin{align}
		\label{eq:dis07}
		p & = i\kappa n\left(e^{i\theta} - 1\right),\nonumber\\
		q & =\theta,\\
		r & = i\kappa n\left(e^{i\theta} - 1 \right),\nonumber\\
		s & = \kappa^2 n^2\left(\frac{e^{i\theta}}{i}-\theta 	-\frac{1}{i}\right)\nonumber.
	\end{align}
	Expanding the exponential $e^{i\theta}$ and keeping the leading order term gives
	\begin{align}
		\label{eq:dis08}
		p & = -\Omega t\kappa n ,\nonumber\\
		q & = \Omega t,\\
		r & = -\Omega t \kappa n,\nonumber\\
		s & = -\frac{\Omega^2 t^2\kappa^2 n^2}{2i}\nonumber.
	\end{align}
	Using (\ref{eq:dis08}) in (\ref{eq:dis03}), we obtain equivalent form of $e^{i\Omega t \left[\frac{\hat{X}^2 + \hat{P}^2}{2} -\sqrt{2}\kappa n\hat{X} \right]}$ that was used in Secs.~\ref{sec:results} and~\ref{sec:comparison}.
	
	%%==========================================================
	\section{Evaluation Of Fourier Coefficient\label{sec:sec:FourierCoeff}}
	%%==========================================================
	The detection of photons is described by the projection operator $\mathcal{P} = \lvert n_c,n_d\rangle\langle n_c,n_d\rvert$. The collapsed
	state of light at the measurement for an nth state of the 	atom is $\langle n\lvert \mathcal{P}\rvert \psi_{BS}(t)\rangle = \langle n_c, n_d, n\lvert \psi_{BS} (t)\rangle\rvert n_c, n_d\rangle$, where $\lvert \psi_{BS} (t) \rangle$ is given in (\ref{eq:strn:cv06}). The probability amplitude $\langle n_c, n_d, n\lvert \psi_{BS} (t)\rangle$ then becomes 
	\begin{align}
		\label{eq:strn:cv07}
		\langle n_c , n_d,n\lvert\psi_{BS}\rangle =  	e^{-\frac{\lvert\alpha_l\rvert^2 +\lvert \alpha_r\rvert^2}{2}}\left(\frac{1}{\sqrt{2}}\right)^{n_c+n_d}\nonumber\\
		\times\sum_{n_l,n_r}\alpha_l^{n_l}\alpha_r^{n_r}
		\delta_{n_l+n_r-n_c,n_d}\nonumber\\
		\times\langle n\rvert e^{ i\Omega t\left[\frac{\hat{X}^2 + 	\hat{P}^2}{2} -\sqrt{2} \kappa (n_l-n_r) \hat{X}\right]  }\lvert0\rangle \nonumber\\
		\times\sum_{k=\mathrm{max}(0,n_l-n_d)}^{\mathrm{min}(n_c,n_l)}\frac{\sqrt{n_c!}\sqrt{n_d!}\, i^{n_l+n_c-2k}}{k!(n_l-k)! (n_c-k)!(n_d-n_l + k)!}
	\end{align}
	In the absence of atom-light interaction, the photon states after recombination at the beamsplitter has a probability amplitude given as~\cite{ilo-okeke2016}
	\begin{align}
		\label{eq:strn23}
		\langle n_c,n_d\lvert\psi\rangle &= 	e^{-\frac{\lvert\alpha_l\rvert^2 +\lvert \alpha_r\rvert^2}{2}}\frac{1}{\sqrt{n_c!}}\frac{1}{\sqrt{n_d!}}\left(\frac{i\alpha_l + \alpha_r}{\sqrt{2}} \right)^{n_c}\times\nonumber\\
		&\left(\frac{\alpha_l + i\alpha_r }{\sqrt{2}}\right)^{n_d}
	\end{align}
	Similarly, (\ref{eq:strn:cv07}) in the same limit ($g=0$) becomes
	\begin{align}
		\label{eq:strn24}
		\langle n_c,n_d\lvert\psi_\mathrm{BS}\rangle & = 	e^{-\frac{\lvert\alpha_l\rvert^2 +\lvert \alpha_r\rvert^2}{2}} \left(\frac{\alpha_r}{\sqrt{2}}\right)^{n_c+n_d} \sqrt{n_c!} \sqrt{n_d!}\nonumber\\
		& \sum_{n_l}\left(\frac{\alpha_l}{\alpha_r}\right)^{n_l} 	S(n_c,n_d,n_l)
	\end{align}
	where 
	%\begin{widetext}
		$$S(n_l,n_c,n_d) =  \sum_{\mathrm{k_{min}}}^{\mathrm{k_{max}}}\frac{ i^{n_l+n_c-2k}}{k!(n_l-k)! (n_c-k)!(n_d-n_l + k)!},$$
		%	\end{widetext} 
	$\mathrm{k_{min}}=\mathrm{max}(0,n_l-n_d)$, and $\mathrm{k_{max}} = \mathrm{min}(n_c,n_l)$. Equations~(\ref{eq:strn23}) and~(\ref{eq:strn24}) though may be functionally different, they represent the same probability amplitude and hence are the same. Equating~(\ref{eq:strn23}) and~(\ref{eq:strn24}), the Fourier transform immediately gives 
	%\begin{widetext}
		\begin{align}
			\label{eq:strn25}
			S(n_l,n_c,n_d) = \left(\frac{\lvert\alpha_l\rvert}{\lvert\alpha_r\rvert}\right)^{-n_l}\frac{1}{n_c!}\frac{1}{n_d!}\frac{1}{2\pi}\times\nonumber\\
			\int_{0}^{2\pi}e^{-in_l\phi}d\phi\left(1 + i\frac{|\alpha_l|}{|\alpha_r|}e^{i\phi} \right)^{n_c}
			\left(i + \frac{|\alpha_l|}{|\alpha_r|}e^{i\phi} \right)^{n_d},
		\end{align}
		%\end{widetext}
	where $\phi = \phi_l - \phi_r$ is the relative phase between $\alpha_l$ and $\alpha_r$. 
	
	From (\ref{eq:strn24}) and (\ref{eq:strn25}) the sum over $k$ in Eq.~(\ref{eq:strn:cv07}) can be replaced and the probability amplitude describing the detection of photons at detectors $c$ and $d$ becomes
	\begin{align}
		\label{eq:strn:cv08}
		\langle n_c, n_d,n\lvert\psi_\mathrm{BS}(t)\rangle  &=  	\frac{e^{-\frac{\lvert\alpha_l\rvert^2 +\lvert \alpha_r\rvert^2}{2}}}{\sqrt{n_c!}\sqrt{n_d!}} \left(\frac{\alpha_r}{\sqrt{2}}\right)^{n_c+n_d}\sum_{n_l}e^{in_l\phi}\nonumber\\
		& I(n_c, n_d, n_l)\\ 
		&\times\langle n\lvert e^{i\Omega t\left(\frac{\hat{X}^2 + 	\hat{P}^2}{2} + \sqrt{2}\kappa (n_c+n_d -2n_l)\hat{X}\right) } \lvert 0 \rangle,\nonumber
	\end{align} 
	where the Fourier coefficient $I(n_c, n_d, n_1)$ is defined in (\ref{eq:strn:cv09a}).
	%	\begin{align}
		%		\label{eq:strn:cv09a}
		%		I(n_1,n_c,n_d) &= \frac{1}{2\pi}\int_{0}^{2\pi}d\phi\, 	e^{-i\,n_1\phi}\left( 1 + i\frac{|\alpha_l|}{|\alpha_r|}e^{i\phi}\right)^{n_c}\times\nonumber\\ &\left(i + \frac{|\alpha_l|}{|\alpha_r|}e^{i\phi} \right)^{n_d},\\
		%		I(n_1,n_c,n_d) &= \frac{1}{2\pi}\int_{-\pi}^{\pi}d\phi\, 	e^{-i\,n_1\phi}\left( 1 + i\frac{|\alpha_l|}{|\alpha_r|}e^{i\phi} \right)^{n_c}\times\nonumber\\
		%		& \left(i + \frac{|\alpha_l|}{|\alpha_r|}e^{i\phi} 	\right)^{n_d}.\nonumber
		%	\end{align}
	%	The interval $\left[0,2\pi\right]$ is convenient when the largest contribution of the integrand comes from points around $\phi \approx \pi$, while the interval $\left[-\pi,\pi\right]$ is convenient when the largest contribution of the integrand comes from points around $\phi \approx 0$. 
	The evaluation of Fourier coefficient (\ref{eq:strn:cv09a})  gives~\cite{ilo-okeke2010,ilo-okeke2016}
	\begin{align}
		\label{eq:strn:cv32}
		I(n_c, n_d, n_1) = \frac{1}{\sqrt{2\pi\sigma}} 	\left[\frac{2n_c}{n_c + n_d}(1+ \eta^2) \right]^{\frac{n_c}{2}}\times\nonumber\\ 
		\left[\frac{2n_d}{n_c + n_d}(1 + \eta^2)\right]^{\frac{n_d}{2}} 	\exp\left[-\frac{(n_c \phi'_{c,0} + n_d\phi'_{d,0} - n_l)^2}{2\sigma}\right]\times\nonumber\\
		\left\{e^{i (n_c\phi_{c,0} + n_d\phi_{d,0} - n_l x_0)} + 	(-1)^{n_d}e^{-i(n_c\phi_{c,0} + n_d\phi_{d,0} + n_l x'_0)} \right\},
	\end{align}
	where $x'_0 = \pi - x_0$, $\eta = \tfrac{\lvert\alpha_l\rvert}{\lvert\alpha_r\rvert}$
	\begin{align}
		\label{eq:strn:cv18}
		x_0& = \arcsin\left(\frac{n_d - n_c}{n_c+ n_d}\frac{1 + 	\eta^2}{2\eta}\right),\\
		\label{eq:strn:cv33}
		\phi_{d,0} & = \arctan\left( \frac{2(n_c + n_d) + (n_d - n_c)(1 + \eta^2)}{\sqrt{4\eta^2(n_c + n_d)^2 - (n_d - n_c)^2(1 + \eta^2)^2}}\right),\\
		\label{eq:strn:cv34}
		\phi_{c,0} & = \arctan\left(\frac{\sqrt{4\eta^2(n_c + n_d)^2 - 	(n_d-n_c)^2(1 + \eta^2)^2}}{2(n_c + n_d ) - (n_d - n_c)(1 + \eta^2)} \right),\\
		\label{eq:strn:cv35}
		\phi'_{d,0} & = \left(\frac{2\eta^2 (n_c+ n_d) + (n_d - n_c)(1 + \eta^2)}{4n_d(1 + \eta^2)}\right),\\
		\label{eq:strn:cv36}
		\phi'_{c,0} & = \left(\frac{2\eta^2(n_c + n_d) - (n_d - n_c)(1 + \eta^2)}{4n_c(1 + \eta^2)} \right).
	\end{align}
	and
	\begin{equation}
		\label{eq:strn:cv28}
		\sigma = \frac{4\eta^2(n_c + n_d)^2 - (n_d - n_c)^2 (1 + 	\eta^2)^2}{8n_cn_d(1 + \eta^2)^2}(n_c + n_d).
	\end{equation}
	
	%%%==========================================================
	\section{Evaluating Photon Probability Amplitude \label{sec:probilityamplitude}}
	%%%==========================================================
	To understand the behavior of photon detection probability distribution, we begin our analysis by evaluation of the probability amplitude (\ref{eq:conditionalstate}). Upon substitution of $I(n_c,n_d,n_l)$, and (\ref{eq:dis03}) in (\ref{eq:conditionalstate}) and summing over the photon input states $n_1$ gives
	\begin{align}
		\label{eq:strn:cv39}
		\langle  n\rvert\psi_{n_c,n_d}\rangle & = 	\frac{e^{-\frac{\lvert\alpha_l\rvert^2 +\lvert\alpha_r\rvert^2}{2}}}{\sqrt{n_c!} \sqrt{n_d!}}\left(\frac{\alpha_r}{\sqrt{2}}\right)^{n_c +n_d}\frac{\left(i\Omega t\kappa\right)^n}{\sqrt{n!}} \nonumber\\ &\times\left(\frac{2n_c(1 + \eta^2)}{n_c + n_d}\right)^{\frac{n_c}{2}} \left(\frac{2n_d(1 + \eta^2)}{n_c + n_d}\right)^{\frac{n_d}{2}}  \nonumber\\&\times \sqrt{\frac{1}{1 + 4\sigma\Omega^2t^2\kappa^2}}\nonumber\\
		&\times\Bigg\{ e^{i[b - a(x_0 	-\phi)]}e^{-\frac{2i\sigma\Omega^2t^2\kappa^2(x_0 - \phi)(n_c + n_d - 2a)}{1+4\sigma\Omega^2t^2\kappa^2}}\nonumber\\
		&\times e^{-\frac{\Omega^2t^2\kappa^2(n_c + n_d 	-2a)^2}{2(1+4\sigma\Omega^2t^2\kappa^2)}} e^{-\frac{\sigma(x_0 -\phi)^2}{2(1+4\sigma\Omega^2t^2\kappa^2)}}\nonumber\\
		&\times\sum_{k=0}^{\frac{n}{2}} 	\frac{n!}{k!(n-2k)!}\left(\frac{2\sigma}{1+ 4\sigma\Omega^2t{^2\kappa^2}}\right)^k\nonumber\\
		&\times\left(n_c+n_d - 2(a + q)\right)^{n-2k}
		+(-1)^{n_c} \nonumber\\
		&\times e^{-i[b + a(x'_0 	-\phi)]}e^{-\frac{2i\sigma\Omega^2t^2\kappa^2(x'_0 - \phi)(n_c + n_d - 2a)}{1+4\sigma\Omega^2t^2\kappa^2}}\nonumber\\
		&\times e^{-\frac{\Omega^2t^2\kappa^2(n_c + n_d 	-2a)^2}{2(1+4\sigma\Omega^2t^2\kappa^2)}} e^{-\frac{\sigma(x'_0 -\phi)^2}{2(1+4\sigma\Omega^2t^2\kappa^2)}}\nonumber\\
		&\times\sum_{k=0}^{\frac{n}{2}} 	\frac{n!}{k!(n-2k)!}\left(\frac{2\sigma}{1+ 4\sigma\Omega^2t{^2\kappa^2}}\right)^k\nonumber\\
		&\times\left(n_c+n_d - 2(a + q')\right)^{n-2k}\Bigg\}
	\end{align}
	where $a = n_c\phi'_{c,0} + n_d\phi'_{d,0}$, and $b = n_c\phi_{c,0}+n_d\phi_{d,0}$, $q = \sigma(2\Omega^2t^2\kappa^2(n_c+n_d -2a) - i(x_0 - \phi))/(1 + 4\sigma\Omega^2t^2\kappa^2)$, and $q'$ is obtained by replacing $x_0$ in $q$ with $x'_0$ (\ref{eq:strn:cv32}). In analysis of the probability amplitude in Sec.~\ref{sec:sec: probden}, the second term of~(\ref{eq:strn:cv39}) is exponentially small for $n_c = n_d$, and   $n_c + n_d -2a =0)$. Hence   the second term is neglected in the analysis of Sec.~\ref{sec:sec: probden}. Notice that the upper limit  in summing over $n$ is fixed by the total number of atoms in the double well trap. 
	
	In Ref.~\cite{ilo-okeke2021}, we decided that we will work within the balanced detection configuration, $\alpha_l = \alpha_r$ and the most probable outcome $n_c,\,n_d = (\lvert\alpha_r|^2 + \lvert\alpha_l\rvert^2)/2$. The justification for these choices becomes obvious since $n_c + n_d - 2a = (n_c+n_d) (\rvert\alpha_r\lvert^2 - \rvert\alpha_l\lvert^2)/(\rvert\alpha_r\lvert^2 + \rvert\alpha_l\lvert^2) = 0$ in the balanced detection configuration. Also, in the regime $\lvert n_c - n_d\rvert \ll n_c + n_d$, $x_0 \approx 0$, and hence $  q = 0$. Thus these conditions combine to limit the back-action of the measurement on quantum state of the atoms. Hence at balanced detection, the probability amplitude simplifies. The amplitude at $x'_0$ is exponentially small and is neglected. Provided that $\frac{2\sigma\Omega^2t^2\kappa^2}{1+ 4\sigma\Omega^2t{^2\kappa^2}}\ll1$, the sum over $n$ converges  rapidly. Taking these into account one obtain the form of the probability density (\ref{eq:papc09}) written in the main text.
	
	%%%==========================================================
	\section{Evaluating Expectations Of Atomic Operators  \label{sec:operators}}
	%%%==========================================================
	Here we evaluate various expectation value of atom operator that are used in Sec.~\ref{sec:sec:ApproximateExpression} for finding the expectations of $\hat{X}$ and  $\hat{P}$, and their variances.  We begin by evaluating the probability amplitude $\langle n\lvert \psi_{n_c,n_d}\rangle$ and expectations of the form $\langle n\lvert A\rvert \psi_{n_c,n_d}\rangle$,  where $A = b^\dagger_e,\, b_e,\, (b^\dagger_e)^2$. Since the $x'_0$ term of the Fourier coefficient~(\ref{eq:strn:cv32}) is exponentially small in the region of interest, it is neglected in the following calculations. Acting the operator $b^\dagger_e$ on the state $\lvert\psi_{n_c,n_d}\rangle$~(\ref{eq:pap03}) and finding its expectation in the atom subspace gives
	\begin{align}
		\label{eq:papc10}
		\langle n\rvert b^\dagger_e\lvert\psi_{n_c,n_d}\rangle & = 	\frac{e^{-\frac{\lvert\alpha_l\rvert^2 +\lvert\alpha_r\rvert^2}{2}}}{\sqrt{n_c!} \sqrt{n_d!}}\left(\alpha_r\right)^{n_c +n_d}\left(\frac{n_c(1 + \eta^2)}{n_c + n_d}\right)^{\frac{n_c}{2}}\nonumber\\ &\times\left(\frac{n_d(1 + \eta^2)}{n_c + n_d}\right)^{\frac{n_d}{2}} \frac{\left(i\Omega t\kappa\right)^{n-1}}{(n-1)!}\nonumber\\
		& \sqrt{\frac{n!}{P(n_c,n_d)}} \sqrt{\frac{1}{1 + 	4\sigma\Omega^2t^2\kappa^2}}\nonumber\\
		&\times\Bigg\{ e^{i[b - a(x_0 	-\phi)]}e^{-\frac{2i\sigma\Omega^2t^2\kappa^2(x_0 - \phi)(n_c + n_d - 2a)}{1+4\sigma\Omega^2t^2\kappa^2}}\nonumber\\
		&\times e^{-\frac{\Omega^2t^2\kappa^2(n_c + n_d 	-2a)^2}{2(1+4\sigma\Omega^2t^2\kappa^2)}} e^{-\frac{\sigma(x_0 -\phi)^2}{2(1+4\sigma\Omega^2t^2\kappa^2)}}\nonumber\\
		&\times\sum_{k=0}^{\frac{n-1}{2}} 	\frac{(n-1)!}{k!(n-1-2k)!}\left(\frac{2\sigma}{1+ 4\sigma\Omega^2t{^2\kappa^2}}\right)^k\nonumber\\
		&\times\left(n_c+n_d - 2(a + q)\right)^{n-1-2k}\Bigg\}
	\end{align}
	where $q = \sigma(2\Omega^2t^2\kappa^2(n_c+n_d -2a) - i(x_0 - \phi))/(1 + 4\sigma\Omega^2t^2\kappa^2)$. For $n_c,n_d = (\lvert\alpha_r\rvert^2 + \lvert\alpha_l\rvert^2)/2$, and $\lvert n_c - n_d\rvert \ll n_c + n_d$ the above simplifies to 
	\begin{align}
		\label{eq:papc11}
		\langle n\rvert b^\dagger_e\lvert\psi_{n_c,n_d}\rangle & = 	\frac{e^{-\frac{\lvert\alpha_l\rvert^2 +\lvert\alpha_r\rvert^2}{2}}}{\sqrt{n_c!} \sqrt{n_d!}}\left(\alpha_r\right)^{n_c +n_d}\left(\frac{n_c(1 + \eta^2)}{n_c + n_d}\right)^{\frac{n_c}{2}}\nonumber\\ &\times\left(\frac{n_d(1 + \eta^2)}{n_c + n_d}\right)^{\frac{n_d}{2}}\sqrt{\frac{1}{P(n_c,n_d)}} \\
		& \times \sqrt{\frac{1}{1 + 	4\sigma\Omega^2t^2\kappa^2}}\left(\frac{2\sigma}{1+ 4\sigma\Omega^2t{^2\kappa^2}}\right)^{\frac{n-1}{2}}\nonumber\\
		&\times e^{ib}\left(i\Omega 	t\kappa\right)^{n-1}\frac{\sqrt{n}!}{\left(\dfrac{n-1}{2}\right)!},\quad\mathrm{for}\, n\,\mathrm{odd},\nonumber\\
		\label{eq:papc12}
		\langle n\rvert b^\dagger_e\lvert\psi_{n_c,n_d}\rangle & = 0, 	\quad \mathrm{for}\, n\,\mathrm{even}.
	\end{align}
	
	Similarly, following same recipe for $b^\dagger_e$, the expectation value of the destruction operator $b_e$ in the atom subspace becomes
	\begin{align}
		\label{eq:papc13}
		\langle n\rvert b_e\lvert\psi_{n_c,n_d}\rangle & = 	\frac{e^{-\frac{\lvert\alpha_l\rvert^2 +\lvert\alpha_r\rvert^2}{2}}}{\sqrt{n_c!} \sqrt{n_d!}}\left(\alpha_r\right)^{n_c +n_d}\left(\frac{n_c(1 + \eta^2)}{n_c + n_d}\right)^{\frac{n_c}{2}}\nonumber\\ &\times\left(\frac{n_d(1 + \eta^2)}{n_c + n_d}\right)^{\frac{n_d}{2}} \frac{\left(i\Omega t\kappa\right)^{n+1}}{\sqrt{n!}}\nonumber\\
		& \sqrt{\frac{1}{P(n_c,n_d)}} \sqrt{\frac{1}{1 + 	4\sigma\Omega^2t^2\kappa^2}}\nonumber\\
		&\times\Bigg\{ e^{i[b - a(x_0 	-\phi)]}e^{-\frac{2i\sigma\Omega^2t^2\kappa^2(x_0 - \phi)(n_c + n_d - 2a)}{1+4\sigma\Omega^2t^2\kappa^2}}\nonumber\\
		&\times e^{-\frac{\Omega^2t^2\kappa^2(n_c + n_d 	-2a)^2}{2(1+4\sigma\Omega^2t^2\kappa^2)}} e^{-\frac{\sigma(x_0 -\phi)^2}{2(1+4\sigma\Omega^2t^2\kappa^2)}}\nonumber\\
		&\times\sum_{k=0}^{\frac{n+1}{2}} 	\frac{(n+1)!}{k!(n+1-2k)!}\left(\frac{2\sigma}{1+ 4\sigma\Omega^2t{^2\kappa^2}}\right)^k\nonumber\\
		&\times\left(n_c+n_d - 2(a + q)\right)^{n+1-2k}\Bigg\}
	\end{align}
	For $n_c,n_d = (\lvert\alpha_r\rvert^2 + \lvert\alpha_l\rvert^2)/2$, and $\lvert n_c - n_d\rvert \ll n_c + n_d$ the above simplifies to 
	\begin{align}
		\label{eq:papc14}
		\langle n\rvert b_e\lvert\psi_{n_c,n_d}\rangle & = 	\frac{e^{-\frac{\lvert\alpha_l\rvert^2 +\lvert\alpha_r\rvert^2}{2}}}{\sqrt{n_c!} \sqrt{n_d!}}\left(\alpha_r\right)^{n_c +n_d}\left(\frac{n_c(1 + \eta^2)}{n_c + n_d}\right)^{\frac{n_c}{2}}\nonumber\\ &\times\left(\frac{n_d(1 + \eta^2)}{n_c + n_d}\right)^{\frac{n_d}{2}}\sqrt{\frac{1}{P(n_c,n_d)}} \\
		& \times \sqrt{\frac{1}{1 + 	4\sigma\Omega^2t^2\kappa^2}}\left(\frac{2\sigma}{1+ 4\sigma\Omega^2t{^2\kappa^2}}\right)^{\frac{n+1}{2}}\nonumber\\
		&\times e^{ib}\frac{\left(i\Omega 	t\kappa\right)^{n+1}}{\sqrt{n!}}\frac{(n+1)!}{\left(\dfrac{n+1}{2}\right)!},\quad\mathrm{for}\, n\,\mathrm{odd},\nonumber\\
		\label{eq:papc15}
		\langle n\rvert b_e\lvert\psi_{n_c,n_d}\rangle & = 0, \quad 	\mathrm{for}\, n\,\mathrm{even}.
	\end{align}
	
	Also, the products of operators that will be used in finding the variance  may be calculated in the same way. For instance, the expectation of operator $(b^\dagger_e)^2$ in the atom subspace is 
	\begin{align}
		\label{eq:papc16}
		\langle n\rvert (b^\dagger_e)^2\lvert\psi_{n_c,n_d}\rangle & = 	\frac{e^{-\frac{\lvert\alpha_l\rvert^2 +\lvert\alpha_r\rvert^2}{2}}}{\sqrt{n_c!} \sqrt{n_d!}}\left(\alpha_r\right)^{n_c +n_d}\left(\frac{n_c(1 + \eta^2)}{n_c + n_d}\right)^{\frac{n_c}{2}}\nonumber\\ &\times\left(\frac{n_d(1 + \eta^2)}{n_c + n_d}\right)^{\frac{n_d}{2}} \frac{\left(i\Omega t\kappa\right)^{n-2}}{(n-2)!}\sqrt{n!}\nonumber\\
		& \sqrt{\frac{1}{P(n_c,n_d)}} \sqrt{\frac{1}{1 + 	4\sigma\Omega^2t^2\kappa^2}}\nonumber\\
		&\times\Bigg\{ e^{i[b - a(x_0 	-\phi)]}e^{-\frac{2i\sigma\Omega^2t^2\kappa^2(x_0 - \phi)(n_c + n_d - 2a)}{1+4\sigma\Omega^2t^2\kappa^2}}\nonumber\\
		&\times e^{-\frac{\Omega^2t^2\kappa^2(n_c + n_d 	-2a)^2}{2(1+4\sigma\Omega^2t^2\kappa^2)}} e^{-\frac{\sigma(x_0 -\phi)^2}{2(1+4\sigma\Omega^2t^2\kappa^2)}}\nonumber\\
		&\times\sum_{k=0}^{\frac{n-2}{2}} 	\frac{(n-2)!}{k!(n-2-2k)!}\left(\frac{2\sigma}{1+ 4\sigma\Omega^2t{^2\kappa^2}}\right)^k\nonumber\\
		&\times\left(n_c+n_d - 2(a + q)\right)^{n-2-2k}\Bigg\}
	\end{align}
	For $n_c,n_d = (\lvert\alpha_r\rvert^2 + \lvert\alpha_l\rvert^2)/2$, $\lvert n_c - n_d\rvert \ll n_c + n_d$ the above simplifies to 
	\begin{align}
		\label{eq:papc17}
		\langle n\rvert (b^\dagger_e)^2\lvert\psi_{n_c,n_d}\rangle & = 	\frac{e^{-\frac{\lvert\alpha_l\rvert^2 +\lvert\alpha_r\rvert^2}{2}}}{\sqrt{n_c!} \sqrt{n_d!}}\left(\alpha_r\right)^{n_c +n_d}\left(\frac{n_c(1 + \eta^2)}{n_c + n_d}\right)^{\frac{n_c}{2}}\nonumber\\ &\times\left(\frac{n_d(1 + \eta^2)}{n_c + n_d}\right)^{\frac{n_d}{2}}\sqrt{\frac{1}{P(n_c,n_d)}} \\
		& \times \sqrt{\frac{1}{1 + 	4\sigma\Omega^2t^2\kappa^2}}\left(\frac{2\sigma}{1+ 4\sigma\Omega^2t{^2\kappa^2}}\right)^{\frac{n-2}{2}}\nonumber\\
		&\times e^{ib}\left(i\Omega 	t\kappa\right)^{n-2}\frac{\sqrt{n!}}{\left(\dfrac{n-2}{2}\right)!},\quad\mathrm{for}\, n\,\mathrm{even},\nonumber\\
		\label{eq:papc18}
		\langle n\rvert (b^\dagger_e)^2\lvert\psi_{n_c,n_d}\rangle & = 0, 	\quad \mathrm{for}\, n\,\mathrm{odd}.
	\end{align}
	The remainder of operators $b^2_e$, $b^\dagger_e b_e$ may calculated following same procedure. Note that the expectations of $(b^\dagger_e)^2$ and $b^2_e$ are equal $\langle \psi_{n_c,n_d}\lvert b_e^2\rvert \psi_{n_c,n_d}\rangle = \langle \psi_{n_c,n_d}\lvert (b^\dagger_e)^2\rvert \psi_{n_c,n_d}\rangle$
	
	Finally for completeness, we calculate the probability amplitude $\langle n\rvert\psi_{n_c,n_d}\rangle $, see also Sec.~\ref{sec:sec: probden}. Projecting the state $\lvert n\rangle$ having $n$ atoms on the state (\ref{eq:pap03}), $\lvert \psi_{n_c,n_d}\rangle$, given that $n_c$ and $n_d$ has been measured  gives
	\begin{align}
		\label{eq:papc19}
		\langle n\rvert\psi_{n_c,n_d}\rangle & = 	\frac{e^{-\frac{\lvert\alpha_l\rvert^2 +\lvert\alpha_r\rvert^2}{2}}}{\sqrt{n_c!} \sqrt{n_d!}}\left(\alpha_r\right)^{n_c +n_d}\left(\frac{n_c(1 + \eta^2)}{n_c + n_d}\right)^{\frac{n_c}{2}}\nonumber\\ &\times\left(\frac{n_d(1 + \eta^2)}{n_c + n_d}\right)^{\frac{n_d}{2}} \frac{\left(i\Omega t\kappa\right)^n}{\sqrt{n!}} \sqrt{\frac{1}{1 + 4\sigma\Omega^2t^2\kappa^2}}\nonumber\\
		&\times\Bigg\{ e^{i[b - a(x_0 	-\phi)]}e^{-\frac{2i\sigma\Omega^2t^2\kappa^2(x_0 - \phi)(n_c + n_d - 2a)}{1+4\sigma\Omega^2t^2\kappa^2}}\nonumber\\
		&\times e^{-\frac{\Omega^2t^2\kappa^2(n_c + n_d 	-2a)^2}{2(1+4\sigma\Omega^2t^2\kappa^2)}} e^{-\frac{\sigma(x_0 -\phi)^2}{2(1+4\sigma\Omega^2t^2\kappa^2)}}\nonumber\\
		&\times\sum_{k=0}^{\frac{n}{2}} 	\frac{n!}{k!(n-2k)!}\left(\frac{2\sigma}{1+ 4\sigma\Omega^2t{^2\kappa^2}}\right)^k\nonumber\\
		&\times\left(n_c+n_d - 2(a + q)\right)^{n-2k}
	\end{align}
	
	For $n_c,n_d = (\lvert\alpha_r\rvert^2 + \lvert\alpha_l\rvert^2)/2$, $\lvert n_c - n_d\rvert\ll n_c + n_d$ the above simplifies to 
	\begin{align}
		\label{eq:papc20}
		\langle n\rvert \psi_{n_c,n_d}\rangle & = 	\frac{e^{-\frac{\lvert\alpha_l\rvert^2 +\lvert\alpha_r\rvert^2}{2}}}{\sqrt{n_c!} \sqrt{n_d!}}\left(\alpha_r\right)^{n_c +n_d}\left(\frac{n_c(1 + \eta^2)}{n_c + n_d}\right)^{\frac{n_c}{2}}\nonumber\\ &\times\left(\frac{n_d(1 + \eta^2)}{n_c + n_d}\right)^{\frac{n_d}{2}}\sqrt{\frac{1}{P(n_c,n_d)}} \\
		& \times \sqrt{\frac{1}{1 + 	4\sigma\Omega^2t^2\kappa^2}}\left(\frac{2\sigma}{1+ 4\sigma\Omega^2t{^2\kappa^2}}\right)^{\frac{n}{2}}\nonumber\\
		&\times e^{ib}\left(i\Omega 	t\kappa\right)^{n}\frac{\sqrt{n!}}{\left(\dfrac{n}{2}\right)!},\quad\mathrm{for}\, n\,\mathrm{even},\nonumber\\
		\label{eq:papc21}
		\langle n\rvert \psi_{n_c,n_d}\rangle & = 0, \quad \mathrm{for}\, 	n\,\mathrm{odd}.
	\end{align}

	\bibliographystyle{apsrev}
%	\bibliography{ReferenceFile}%The following command as it is allows me to have a central .bib file which I can always update
%	\bibliography{C:/Users/Phymbas/Documents/Bibiliography/ReferenceFile}

	%
	%
\end{document}